\documentclass[aps,pra,showpacs,twocolumn,floatfix]{revtex4}
\usepackage{amsmath}
\begin{document}

\title{Relativistic many-body calculation of energies, oscillator strengths, transition rates,
 lifetimes,  polarizabilities, and quadrupole moment of Fr-like Th~IV  ion}

\author{ M. S. Safronova$^{1,2}$ and  U. I. Safronova$^{3,4}$ }

\affiliation {$^1$Department of Physics and Astronomy, University of Delaware, Newark, Delaware 19716\\
$^2$Joint Quantum Institute, NIST and the University of Maryland, College Park, Maryland 20742\\
$^3$Physics Department, University of Nevada, Reno, Nevada 89557,
\\$^4$Department of Physics,  University of Notre Dame,
Notre Dame, IN 46556}

\date{\today}

\begin{abstract}
Atomic properties of the 24 low-lying $ns$, $np_j$, $nd_j$, $nf_j$,
and $ng_j$ states in Th~IV ion are calculated using the
high-precision relativistic all-order
 method where all single, double, and partial triple excitations
of the Dirac-Fock wave functions are included to all orders of perturbation theory. Recommended values are provided for a
large number of electric-dipole matrix elements,
oscillator strengths, transition rates, and lifetimes. Scalar polarizabilities of the ground and six excited
states ($5f_j$, $6d_j$, $7p_j$, and $7s$ ), and tensor
polarizabilities of the $5f_j$, $6d_j$, $7p_{3/2}$ states of Th~IV are evaluated.
The uncertainties of the recommended values are estimated.
 These calculations provide recommended values critically evaluated
for their accuracy for a number of Th~IV atomic properties for use in theoretical modeling as well as planning and
analysis of various experiments including development of ultra precise nuclear clock and RESIS studies of actinide
ions.

%31.15.A- Ab initio calculations
%31.15.ac High-precision calculations for few-electron (or few-body) atomic systems
%31.15.ag Excitation energies and lifetimes; oscillator strengths
%31.15.aj Relativistic corrections, spin-orbit effects, fine structure; hyperfine structure
%31.15.am Relativistic configuration interaction (CI) and many-body perturbation calculations
%31.15.ap Polarizabilities and other atomic and molecular properties
%31.15.V- Electron correlation calculations for atoms, ions and molecules
%31.15.vj Electron correlation calculations for atoms and ions: excited states
%31.30.J- Relativistic and quantum electrodynamic (QED) effects in atoms, molecules, and ions
\end{abstract}
\maketitle

%\end{document}

%*************************
\section{Introduction}
The $^{229}$Th nuclear excitation energy of a few eV ~\cite{1,2} presents remarkable opportunity to develop an
ultraprecise clock bases on this very narrow nuclear transition ~\cite{peik,gt2,CamRadKuz12}. This transition was also
proposed ~\cite{5} for the laboratory search for variation of the fine structure constant and the dimensionless strong
interaction parameter $m_q/\Lambda_{\textrm{QCD}}$ due to estimated 5-6 orders of magnitude enhancement.
  The
energy splittings of the ground and excited states of the nuclei
are generally much larger and are not accessible with laser
spectroscopy.
 In 2009, laser cooling of the $^{229}$Th$^{3+}$ was reported by
Campbell {\textit{et al.}}  \cite{gt2}. This was the first
demonstration of laser cooling of a multiply-charged ion.
Laser-cooled Wigner crystals $^{229}$Th$^{3+}$ were produce in a
linear Paul trap ~\cite{gt2}. These experimental advances opened
an avenue for excitation of the nuclear transition in a trapped,
cold $^{229}$Th$^{3+}$ ion that may lead to a new levels of
metrological precision \cite{gt2,gt3,CamRadKuz12}.

While the clock based of ultraviolet  $^{229}$Th nuclear transition can be designed with various Th
ions, Th$^{3+}$ is particularly attractive due to its simples electronic structure of one valence
electron above the closed [Rn]=[Xe]$4f^{14}5d^{10}6s^26p^6$ core. The transition probability of the
Th$^{229}$ nucleus from its lowest-energy isomeric states  to the ground state due to the
electronic bridge process was evaluated in  \cite{eb}. Implementation of the electronic bridge
process will require good understanding of Th$^{3+}$ atomic properties, including matrix elements
of the electric-dipole and hyperfine operators. A single-ion nuclear clock based of the stretched
states within the $5f_{5/2}$ electronic ground states of both nuclear isomeric and ground manifolds
was recently proposed in ~\cite{CamRadKuz12}.

 The hyperfine A and B constants for the $5f_{5/2}$,
$5f_{7/2}$, $6d_{3/2}$, and $6d_{5/2}$ states were recently measured allowing to determine nuclear electric-quadrupole
moment $Q=3.11(16)$~eb ~\cite{gt3}. Accurate measurement of the hyperfine constants, combined with precision
theoretical calculations may be used to produce more accurate determination of the $^{229}$Th nuclear magnetic moment,
which is presently known to about 10\%~\cite{nuc}. The relative isotope shifts with respect to $^{232}$Th$^{3+}$ were
measured for three $5f-6d$ transitions ~\cite{gt3}. The 717-nm electric quadrupole transition was observed in
\cite{RadCamKuz12}; the $6d_{3/2}-7s$ transition frequency and the lifetime of the metastable $7s$ level were measured
to be 417845964(30)~MHz and 0.60(7)~s, respectively.

In 2011, binding energies of high-L Rydberg states ($L\geq 7$) of Th$^{2+}$ with $n = 27-29$ were studied using the
resonant excitation Stark ionization spectroscopy (RESIS) method ~\cite{lundeen}. Analysis of the observed RESIS
spectra led to determination of five properties of the Th$^{3+}$ ion: its electric quadrupole moment, adiabatic scalar
and tensor dipole polarizabilities, and the dipole matrix elements connecting the ground $5f_{5/2}$ level to the
low-lying $6d_{3/2}$ and $6d_{5/2}$ levels.

The optical spectroscopy has been reported for Th$^{3+}$
\cite{expt-en}, determining the relative positions of the lowest
24 levels, but most Th$^{3+}$ properties of even low-lying levels
are not known experimentally.

 Recently, oscillator strengths and transition rates were
reported by Safronova {\it et al.\/} \cite{safr-07-fr,safr-06-th},
Migdalek {\it et al.\/} \cite{migd-07}, and Bi\'{e}mont {\it et
al.\/} \cite{osc-ra}. The pseudo-relativistic Hartree-Fock
(HFR+CP) method including core-polarization effects was used in
Ref.~\cite{osc-ra} to evaluate oscillator strengths and transition
rates for the 76 transitions in Th$^{3+}$ ion ~\cite{osc-ra}.
Dirac-Fock + core-polarization approximation, where core-valence
electron correlations were treated in a semiclassical
core-polarization picture, was used to evaluate properties of 20
E1 transitions in Th~IV in Ref.~\cite{migd-07}. Excitation
energies, reduced matrix elements, oscillator strengths,
transition rates,  scalar and tensor ground states
polarizabilities, and lifetimes for a large number of levels  were
calculated  in \cite{safr-06-th} using the third-order many-body
perturbation theory (MBPT) and single-double (SD) all-order
methods.

Accurate calculations of Th$^{3+}$ atomic properties are difficult. While it is a Fr-like ion, its level structure is
different from both Fr and Ra$^+$, which both have $7s$ ground state. Th$^{3+}$ ground state is $5f_{5/2}$ causing
further difficulties in the accurate calculation of its properties not present in either Fr or Ra$^+$. Moreover,
Th$^{3+}$ is sufficiently multicharged to make Breit contributions significant. Most of theoretical and experimental
high-precision studies involved $ns$, $np$, and $nd$ levels resulting in lack of  benchmarks for $nf$ state properties
in other systems that may be used to further improve \textit{ab initio} calculations. The study of the electronic
bridge process ~\cite{eb} noted rather poor agreement of theoretical and experimental energies.

Therefore, we calculate properties of Th$^{3+}$ ion by several different approaches to study the correlation
contributions to various properties to evaluate accuracy of our calculations and to provide a pathway to further
improvement in theoretical understanding of this ion. Due to above noted interesting applications, Th$^{3+}$ also
represent an excellent benchmark for further development of high-precision methodologies of very heavy ions.

 In the present work, we evaluated all properties using both SD and single-double partial triple (SDpT)
 all-order methods as well as carried out additional scaling to evaluate dominate  missing correlation corrections
 and evaluate uncertainties of our calculations. The SD and SDpT methods and their application were discussed in a review
 \cite{review07} and references therein.
Energies and lifetimes are calculated for the $ns$ ($n = 7-10$), $np$ ($n = 7-8$), $nd$ ($n = 6-8$), $nf$ ($n = 5-7$),
and $ng$ ($n = 5-6$) states.
  Reduced matrix
elements, oscillator strengths, and transition rates  are calculated for allowed electric-dipole transitions between
these states. Scalar polarizabilities of the seven first $5f_j$, $6d_j$, $7p_j$, and $7s$ states, and tensor
polarizabilities of the $5f_j$, $6d_j$, and $7p_{3/2}$ states of Th$^{3+}$ are evaluated.
 Particular care was taken to
accurately treat contributions from  highly-excited states.  The present calculation of the transition rates,
lifetimes, and polarizabilities required accurate representation of rather highly excited states, such as $7l_j$,
leading to the use of the large R = 100 a.u. cavity for the generation of the finite B-spline basis set \cite{Bspline}
and higher number of splines N=70 to produce high-accuracy single-particle orbitals. The methods for evaluating the
uncertainties of theoretical values calculated in the framework of the all-order approach are discussed. The
calculation of uncertainties involved estimation of missing high-order effects and {\it ab initio} calculations in
different approximations to establish the size of the higher-order corrections and to approximate missing
contributions.

\section{Energy levels and transition properties}
\subsection{Energy levels}
\begin{table*}
\caption{\label{tab-dip} Recommended values of the reduced
electric-dipole
 matrix elements in Th~IV in atomic units.
 The
 first-order, second-order,  third-order MBPT, and all-order SD and SDpT  values are listed;
 the label ``sc'' indicates the scaled values. Absolute values are given. Final recommended values and their
uncertainties are given in the $Z^\text{{final}}$ column. The last
column gives relative uncertainties of the final values in \%. }
\begin{ruledtabular}
\begin{tabular}{llrrrrrrrrc}
\multicolumn{2}{c}{Transition}& \multicolumn{1}{c}{$Z^{{\rm DF}}$
}& \multicolumn{1}{c}{$Z^{({\rm DF}+2)}$ }&
\multicolumn{1}{c}{$Z^{({\rm DF}+2+3)}$}&
\multicolumn{1}{c}{$Z^\text{{SD}}$ }& \multicolumn{1}{c}{$Z_{\rm
sc}^\text{{(SD)}}$ }& \multicolumn{1}{c}{$Z^\text{{SDpT}}$ }&
\multicolumn{1}{c}{$Z_{\rm sc}^\text{{SDpT}}$ }&
\multicolumn{1}{c}{$Z^\text{{final}}$ }&
\multicolumn{1}{c}{Unc. (\%)}\\
\hline
   $7p_{1/2}$&$   6d_{3/2}$&   2.5465&     2.1960&      2.0566&    2.1220&     2.1284&     2.1312&    2.1253&     2.122(30)&     1.4 \\
   $7p_{1/2}$&$   7d_{3/2}$&   3.8261&     3.5613&      3.4020&    3.4490&     3.4642&     3.4635&    3.4638&     3.449(26)&     0.8 \\
   $7p_{1/2}$&$   7s_{1/2}$&   2.8994&     2.4738&      2.3476&    2.4197&     2.4368&     2.4323&    2.4364&     2.420(34)&     1.4 \\
   $7p_{1/2}$&$   8s_{1/2}$&   1.5874&     1.6297&      1.5549&    1.5492&     1.5404&     1.5542&    1.5390&     1.549(19)&     1.3 \\
   $7p_{1/2}$&$   9s_{1/2}$&   0.4722&     0.4910&      0.4734&    0.4657&     0.4630&     0.4667&    0.4641&     0.463(5) &     1.1 \\
   $7p_{3/2}$&$   6d_{3/2}$&   0.9963&     0.8823&      0.8260&    0.8488&     0.8516&     0.8533&    0.8501&     0.849(10)&     1.2 \\
   $7p_{3/2}$&$   6d_{5/2}$&   3.1975&     2.8762&      2.6900&    2.7550&     2.7627&     2.7665&    2.7583&     2.755(31)&     1.1 \\
   $7p_{3/2}$&$   7d_{3/2}$&   2.0308&     1.8920&      1.8252&    1.8445&     1.8522&     1.8503&    1.8523&     1.845(13)&     0.7 \\
   $7p_{3/2}$&$   7d_{5/2}$&   5.9481&     5.5572&      5.3603&    5.4192&     5.4409&     5.4377&    5.4415&     5.419(37)&     0.7 \\
   $7p_{3/2}$&$   7s_{1/2}$&   3.9933&     3.4515&      3.2731&    3.3677&     3.3925&     3.3866&    3.3919&     3.368(44)&     1.3 \\
   $7p_{3/2}$&$   8s_{1/2}$&   3.0768&     3.0702&      2.9863&    2.9756&     2.9635&     2.9804&    2.9623&     2.963(24)&     0.8\\
   $7p_{3/2}$&$   9s_{1/2}$&   0.7567&     0.7439&      0.7157&    0.7123&     0.7197&     0.7141&    0.7216&     0.712(10)&     1.5 \\
   $8p_{1/2}$&$   7d_{3/2}$&   5.4788&     5.3607&      5.1639&    5.1791&     5.1866&     5.1907&    5.1838&     5.179(11)&     0.2\\
   $8p_{1/2}$&$   8s_{1/2}$&   5.0325&     4.8413&      4.6814&    4.7280&     4.7590&     4.7405&    4.7579&     4.728(21)&     0.5 \\
   $8p_{3/2}$&$   7d_{3/2}$&   2.1716&     2.1485&      2.0566&    2.0630&     2.0641&     2.0690&    2.0628&     2.064(05)&     0.2\\
   $8p_{3/2}$&$   7d_{5/2}$&   6.8642&     6.7805&      6.5104&    6.5180&     6.5247&     6.5343&    6.5209&     6.525(10)&     0.2\\
   $8p_{3/2}$&$   8s_{1/2}$&   6.7737&     6.5493&      6.3192&    6.3881&     6.4284&     6.4068&    6.4270&     6.388(57)&     0.9\\
   $8p_{3/2}$&$   9s_{1/2}$&   5.4177&     5.4364&      5.3083&    5.2854&     5.2662&     5.2961&    5.2631&     5.266(30)&     0.6\\
   $5f_{5/2}$&$   6d_{3/2}$&   2.4281&     1.6597&      1.3609&    1.5296&     1.5330&     1.5423&    1.5231&     1.530(63)&     4.1 \\
   $5f_{5/2}$&$   6d_{5/2}$&   0.6391&     0.4586&      0.3685&    0.4116&     0.4125&     0.4154&    0.4100&     0.412(16)&     3.9 \\
   $5f_{5/2}$&$   5g_{7/2}$&   1.1236&     0.8404&      0.6123&    0.6895&     0.6544&     0.7034&    0.6555&     0.690(30)&     4.4 \\
   $5f_{5/2}$&$   7d_{3/2}$&   0.0654&     0.3417&      0.2077&    0.2588&     0.2475&     0.2710&    0.2449&     0.259(26)&     9.9 \\
   $5f_{5/2}$&$   7d_{5/2}$&   0.0048&     0.0671&      0.0362&    0.0521&     0.0492&     0.0554&    0.0486&     0.052(8) &    15 \\
   $5f_{7/2}$&$   6d_{5/2}$&   2.9557&     2.1257&      1.7270&    1.9191&     1.9223&     1.9371&    1.9122&     1.919(73)&     3.8 \\
   $5f_{7/2}$&$   5g_{7/2}$&   0.2298&     0.1760&      0.1292&    0.1478&     0.1407&     0.1506&    0.1408&     0.148(6) &     3.9 \\
   $5f_{7/2}$&$   5g_{9/2}$&   1.3635&     1.0667&      0.7894&    0.8855&     0.8487&     0.9022&    0.8484&     0.885(33)&     3.8 \\
   $5f_{7/2}$&$   7d_{5/2}$&   0.0703&     0.3889&      0.2259&    0.2961&     0.2815&     0.3114&    0.2785&     0.296(35)&    12 \\
   $6f_{5/2}$&$   6d_{3/2}$&   2.6761&     2.3276&      2.3423&    2.3443&     2.3181&     2.3372&    2.3220&     2.344(23)&     1.0 \\
   $6f_{5/2}$&$   6d_{5/2}$&   0.7669&     0.6780&      0.6837&    0.6800&     0.6727&     0.6774&    0.6736&     0.680(6) &     0.9 \\
   $6f_{7/2}$&$   6d_{5/2}$&   3.3539&     2.9806&      3.0233&    3.0008&     2.9678&     2.9886&    2.9709&     3.001(25)&     0.8 \\
   $7f_{5/2}$&$   8d_{3/2}$&  13.4659&    13.2664&     12.6363&   12.5835&    12.7110&    12.6315&   12.6803&     12.71(13)&     1.0 \\
   $7f_{5/2}$&$   8d_{5/2}$&   3.5961&     3.5454&      3.3681&    3.3554&     3.3900&     3.3553&    3.3662&     3.390(35)&     1.0 \\
   $7f_{5/2}$&$   7d_{5/2}$&   1.0359&     1.0451&      1.1332&    1.1178&     1.0853&     1.1014&    1.0845&     1.085(16)&     1.5 \\
   $7f_{5/2}$&$   6d_{3/2}$&   1.2888&     0.9357&      0.8390&    0.8131&     0.8224&     0.8260&    0.8297&     0.813(33)&     4.1 \\
   $7f_{5/2}$&$   6d_{5/2}$&   0.3586&     0.2713&      0.2376&    0.2286&     0.2321&     0.2323&    0.2340&     0.229(09)&     4.0 \\
   $7f_{5/2}$&$   5g_{7/2}$&   8.5105&     8.5592&      8.6347&    8.3981&     8.4526&     8.4008&    8.4515&     8.453(52)&     0.6\\
   $7f_{7/2}$&$   5g_{9/2}$&   9.5778&     9.6340&      9.7290&    9.4935&     9.4490&     9.4944&    9.4466&     9.449(45)&     0.5\\
   $7f_{7/2}$&$   5g_{7/2}$&   1.6217&     1.6312&      1.6446&    1.5971&     1.5952&     1.5988&    1.5950&     1.595(4)&      0.2\\
   $7f_{7/2}$&$   6d_{5/2}$&   1.5854&     1.2543&      1.1202&    1.0734&     1.0900&     1.0872&    1.0966&     1.073(36)&     3.3 \\
\end{tabular}
\end{ruledtabular}
\end{table*}
 The calculation of energies in Th$^{3+}$ was discussed in detail by Safronova {\it et al.\/} \cite{safr-06-th}
where the third-order relativistic many-body perturbation theory (RMBPT) and all-order SD energies were presented. The
third-order RMBPT approximation includes the second-order and third-order part of the correlation energies. The
all-order  SD approximation includes the second-order and the single-double part of the higher-order correlation
energies. However, it is missing the part of the third-order contribution $E^{(3)}_\text{{extra}}$. The additional
third-order contribution to the energies was added in \cite{safr-06-th} using a separate calculation. The inclusion of
the partial triple-excitations terms via the SDpT method described in \cite{review07} and implemented in the present
work automatically includes the missing third-order energy. The data in \cite{safr-06-th} show extremely large
contributions of the correlation corrections into the energy values. In fact, the lowest-order Dirac-Fock calculation
gives $6d_{3/2}$ ground state instead of the $5f_{5/2}$ state. We find that triple excitations beyond the third-order
term $E^{(3)}_\text{{extra}}$ are very large, 3-5\% of the total correlation correction. For example, the difference of
the correlation correction to the ionization potential calculated in the SD approximation + $E^{(3)}_\text{{extra}}$
term and the SDpT value is 1200~cm$^{-1}$. Based on the size of all other corrections and experimental values, we
estimate that the omitted triple and higher effects for the $5f$ and $6d$ state are on the order of already included
triple excitations, which is reasonable expectation of the accuracy in this case.  The relative contribution of the
correlations is substantially higher (by at least a factor of two) for the $5f$ states than for all other states
exacerbating the problem for the  transition energies. Therefore, full inclusion of the triple excitations, and most
likely an estimate of the higher excitations would be required for accurate description of the energy levels
differences with the ground state owing to significant imbalance of the correlation contribution between the ground
state and all other states except $5f_{7/2}$. Our values for the fine structure $5f_{5/2}-5f_{7/2}$ interval,
4165~cm$^{-1}$ is in good agreement with experiment, 4325~cm$^{-1}$. We included the Breit interaction on the same
footing as the Coulomb interaction in the basis set, which incorporates high-order Breit effects. The Breit interaction
was included to second order in \cite{safr-06-th}, which significantly overestimates its correction. We also note that
inclusion of the higher partial waves with $l>6$ is very important for accurate description of the $5f$ states. The
contribution of $l>6$ is on the order of $1000$~cm$^{-1}$ for the $5f$ states and $250-300$~cm$^{-1}$ for the $6d$
states. We use experimental energy intervals in calculation of all transitions properties and polarizabilities below,
where available.
\subsection{Electric-dipole matrix elements}
\label{E1} In Table \ref{tab-dip}, we list our recommended values for E1 $n'p-ns$, $nd-n'p$,  $nd-n'f$, and $ng-n'f$
transitions. We note that we have calculated about 80 E1 matrix elements to consider all dipole transitions between
$ns$, $np$, $n'd$, $n''f$, and $n''g$ states with $n= 7-10$, $n' = 6-8$, and $n'' = 5-7$. We refer to these values as
the ``best set'' of the matrix elements. We list only the matrix elements that give significant contributions to the
atomic properties calculated in the other sections. To evaluate the uncertainties of these values, we carried out
several calculations in different approximations. To demonstrate the size of the second, third, and higher-order
correlation corrections, we list the lowest-order Dirac-Fock (DF) $Z^{{\rm DF}}$, second-order $Z^{({\rm DF}+2)}$, and
third-order $Z^{({\rm DF}+2+3)}$ values in the first three numerical columns of Table~\ref{tab-dip}. The absolute
values in atomic units ($a_0e$) are given in all cases. The third-order MBPT calculations are carried out following the
method described in Ref.~\cite{adndt-96}.
 The   $Z^{({\rm DF}+2)}$ values are obtained as the sum of
 the second-order correlation correction $Z^{(2)}$
   and  the DF matrix elements $Z^{\rm DF}$.
 The third-order matrix elements $Z^{({\rm DF}+2+3)}$ include the DF values,
 the second-order $Z^{(2)}$ results,
and the third-order $Z^{(3)}$  correlation correction. $Z^{(3)}$ includes random-phase-approximation  terms (RPA)
iterated  to all orders, Brueckner orbital (BO) corrections, the structural radiation,  and normalization terms (see
\cite{adndt-96} for definition of these terms). Next four columns give the results of four different all-order
calculations. \textit{Ab initio} electric-dipole matrix elements evaluated in the all-order  SD (single-double)  and
SDpT approximations (single-double all-order method including partial triple excitations \cite{mar-pol-99}) are given
in columns labeled $Z^\text{SD}$ and $Z^\text{SDpT}$
 of Table~\ref{tab-dip}. The SD and SDpT matrix elements
$Z^\text{SD}$ include $Z^{(3)}$ completely, along with important fourth- and higher-order corrections.  Difference
between the $Z^\text{SD}$ and $Z^\text{SDpT}$ values is about 0.2~\% - 2~\%, i.e. the effect of the triple excitations
on the values of matrix elements is significantly smaller than for the energies.

 The last column of Table~\ref{tab-dip} gives relative uncertainties
of the final values $Z^{\rm final}$ in \%. We use two different methods for the estimation of the uncertainties based
on the type of the dominant correlation corrections for a specific transition. If the correlation terms containing
valence single-excitation coefficients are dominant, the omitted correlation corrections can be estimated by a scaling
procedure described, for example, in Ref.~\cite{SafSaf11rb}.
 In
this case, we use well-defined and rather accurate procedure for the  evaluation of the uncertainty of the matrix
elements described in detail in \cite{SafSaf11rb,SafSaf12,SafSaf13}. It is based on four different
 all-order calculations that included two \textit{ab initio} all-order calculations with (SDpT)
  and without (SD) the inclusion of the partial triple excitations and two
calculations that included semiempirical estimate of high-order correlation corrections starting from both \textit{ab
initio} runs, SD$_{sc}$ and SDpT$_{sc}$. The differences of these
 four values were used to  estimate uncertainty in the final
result  for each transition and
 the SD scaled values are taken as final for these cases.
We note that the scaling may be less reliable in Th$^{3+}$ than in other systems due to large uncertainty in the
experiment ionization potential 231065(200)~cm$^{-1}$~\cite{RalKraRea11} as scaling relies on the experimental values
of the removal energies. However, \textit{ab initio} SDpT results are generally already close to the final scaled
values.

Unfortunately, different type of the correlation terms is dominant for a large fraction of the  transitions of interest
for this work (including all of the transitions containing the $5f$ states). In these case, the above strategy for
evaluating uncertainties is expected to underestimate the uncertainties. We have developed a different approach for
these cases using the study of uncertainties in a similar reference ion, Rb-like Y, where the above (scaling) procedure
is expected to work well \cite{SafSaf13}. We have compared the estimated uncertainties for 60 $nd-n^{\prime}f$
transitions \cite{SafSaf13} with the size of the correlation corrections for the same transitions. We find that
\textit{on average}, the estimate uncertainty was about 7\% of the correlation correction, which was calculated as the
difference of the all-order and the lowest-order results. Therefore, we use 7\% of the correlation correction to
estimate the uncertainties as the second method for evaluating the uncertainties and list these uncertainties for
transitions where the first method is not expected to produce reliable results. We note that the second method is less
precise that the first one and provides a rough estimate of the accuracy. The \textit{ab initio} SD data are listed as
final for these transitions. An accurate benchmark reference measurement is needed to improve the accuracy estimates.

We find three cases in Table~\ref{tab-dip} where neither of the two methods is expected to provide accurate estimates
of the uncertainties. For all three $5f-7d$ transitions, the lowest order values are less than 0.1~a.u. and almost
entire values come from the correlation correction. In these cases, we took 50\%  of the entire higher-order correction
(calculated as the difference of the all-order and the third-order values) as the uncertainty. The 50\% was chosen
based on the comparison of the higher-order effects for other transitions with the corresponding estimates of their
uncertainties carried out by the other methods.
\begin{table*}
\caption{\label{tab-tran} Wavelengths $\lambda$ (\AA) and
transition rates $A_r$ (s$^{-1}$)  for transitions in Th~IV
calculated using our recommended values of reduced electric-dipole
matrix elements
 $A_{r}^\text{{final}}$ and
their uncertainties.  The relative uncertainties are listed in column ``Unc.'' in \%.  In columns $\lambda$, we list
experimental data  \cite{expt-en}. Numbers in brackets represent powers of 10.}
\begin{ruledtabular}
\begin{tabular}{llclcllllcllllc}
\multicolumn{2}{c}{Transition} & \multicolumn{1}{c}{$\lambda$} &
\multicolumn{1}{c}{$A_r$} & \multicolumn{1}{c}{Unc.}&
\multicolumn{2}{c}{Transition} & \multicolumn{1}{c}{$\lambda$} &
\multicolumn{1}{c}{$A_r$} & \multicolumn{1}{c}{Unc.}&
\multicolumn{2}{c}{Transition} & \multicolumn{1}{c}{$\lambda$} &
\multicolumn{1}{c}{$A_r$} &
\multicolumn{1}{c}{Unc.} \\
\multicolumn{1}{c}{lower} & \multicolumn{1}{c}{upper} &
\multicolumn{1}{c}{\AA} & \multicolumn{1}{c}{s$^{-1}$} &
\multicolumn{1}{c}{\%} & \multicolumn{1}{c}{lower} &
\multicolumn{1}{c}{upper} & \multicolumn{1}{c}{\AA} &
\multicolumn{1}{c}{s$^{-1}$} & \multicolumn{1}{c}{\%} &
\multicolumn{1}{c}{lower} & \multicolumn{1}{c}{upper} &
\multicolumn{1}{c}{\AA} & \multicolumn{1}{c}{s$^{-1}$} &
\multicolumn{1}{c}{\%}\\
\hline
  $ 5f_{5/2}$&$  8d_{5/2}$&     615.28&  3.08[6]&    12& $ 7p_{3/2}$&$  9s_{1/2}$&    1140.61&   3.46[8]&   3.0&  $ 8p_{3/2}$&$  8d_{3/2}$&    4528.35&   5.21[7]&    0.5\\
  $ 5f_{5/2}$&$  8d_{3/2}$&     617.46&  9.47[7]&    14& $ 6d_{3/2}$&$  7p_{3/2}$&    1565.86&   9.51[7]&   2.4&  $ 8p_{3/2}$&$  9s_{1/2}$&    4794.51&   2.55[8]&    1.1\\
  $ 5f_{5/2}$&$  5g_{7/2}$&     627.39&  4.88[8]&    8.8& $ 7p_{1/2}$&$  7d_{3/2}$&    1682.21&   1.27[9]&   1.5&  $ 8s_{1/2}$&$  8p_{3/2}$&    4938.44&   1.72[8]&    1.8\\
  $ 5f_{7/2}$&$  8d_{5/2}$&     632.10&  8.49[7]&    14& $ 7p_{1/2}$&$  8s_{1/2}$&    1684.00&   5.09[8]&   2.5&  $ 7d_{3/2}$&$  8p_{3/2}$&    4953.85&   1.78[7]&    0.5\\
  $ 6d_{3/2}$&$  7f_{5/2}$&     643.66&  8.37[8]&    8.2& $ 6d_{5/2}$&$  7p_{3/2}$&    1707.37&   7.72[8]&   2.2&  $ 7d_{5/2}$&$  8p_{3/2}$&    5421.88&   1.35[8]&    0.3\\
  $ 5f_{7/2}$&$  5g_{7/2}$&     644.89&  2.06[7]&    7.8& $ 6d_{3/2}$&$  7p_{1/2}$&    1959.02&   6.07[8]&   2.8&  $ 7f_{5/2}$&$  6g_{7/2}$&    5841.02&   2.22[8]&    5.3\\
  $ 5f_{7/2}$&$  5g_{9/2}$&     644.97&  5.92[8]&    7.6& $ 7s_{1/2}$&$  7p_{3/2}$&    2003.00&   7.15[8]&   2.6&  $ 7f_{7/2}$&$  6g_{7/2}$&    6018.30&   7.90[6]&    3.0\\
  $ 6d_{5/2}$&$  7f_{7/2}$&     664.13&  9.96[8]&    6.7& $ 7p_{3/2}$&$  7d_{5/2}$&    2067.35&   1.12[9]&   1.4&  $ 7f_{7/2}$&$  6g_{9/2}$&    6018.66&   2.21[8]&    2.6\\
  $ 6d_{5/2}$&$  7f_{5/2}$&     666.36&  5.96[7]&    8.0& $ 8p_{1/2}$&$ 10s_{1/2}$&    2086.62&   6.81[7]&   2.7&  $ 8s_{1/2}$&$  8p_{1/2}$&    6713.71&   7.48[7]&    0.9\\
  $ 6d_{3/2}$&$  8p_{3/2}$&     765.24&  2.29[7]&    7.4& $ 7p_{3/2}$&$  7d_{3/2}$&    2144.60&   1.75[8]&   1.4&  $ 7d_{3/2}$&$  8p_{1/2}$&    6742.22&   8.87[7]&    0.4\\
  $ 6d_{5/2}$&$  8p_{3/2}$&     797.55&  1.84[8]&    7.3& $ 7p_{3/2}$&$  8s_{1/2}$&    2147.50&   8.98[8]&   1.6&  $ 5f_{5/2}$&$  6d_{5/2}$&    6903.05&   1.74[5]&    7.7\\
  $ 6d_{3/2}$&$  8p_{1/2}$&     797.94&  7.24[7]&    16& $ 7d_{3/2}$&$  7f_{5/2}$&    2228.66&   4.22[8]&   3.4&  $ 5f_{7/2}$&$  6d_{5/2}$&    9841.58&   1.30[6]&    7.6\\
  $ 7p_{1/2}$&$ 10s_{1/2}$&     818.32&  1.14[8]&    0.7& $ 7d_{5/2}$&$  7f_{7/2}$&    2291.90&   4.51[8]&   7.2&  $ 5f_{5/2}$&$  6d_{3/2}$&   10877.60&   9.21[5]&    8.2\\
  $ 6d_{3/2}$&$  6f_{5/2}$&     846.91&  3.06[9]&    2.0& $ 7d_{5/2}$&$  7f_{5/2}$&    2318.70&   3.19[7]&   3.0&  $ 7d_{3/2}$&$  6f_{5/2}$&   13184.61&   7.34[6]&    3.0\\
  $ 6d_{5/2}$&$  6f_{7/2}$&     882.39&  3.32[9]&    1.6& $ 8p_{3/2}$&$ 10s_{1/2}$&    2349.07&   1.08[8]&   1.4&  $ 7d_{5/2}$&$  6f_{7/2}$&   15653.86&   4.76[6]&    3.0\\
  $ 6d_{5/2}$&$  6f_{5/2}$&     886.66&  2.24[8]&    1.8& $ 7s_{1/2}$&$  7p_{1/2}$&    2694.81&   3.03[8]&   2.8&  $ 7d_{5/2}$&$  6f_{5/2}$&   17117.13&   2.38[5]&    3.1\\
  $ 7s_{1/2}$&$  8p_{1/2}$&     897.78&  7.71[7]&    11& $ 6f_{5/2}$&$  5g_{7/2}$&    3113.25&   6.75[8]&   3.0&  $ 5g_{9/2}$&$  7f_{7/2}$&   17580.87&   4.16[6]&    1.0\\
  $ 7p_{3/2}$&$ 10s_{1/2}$&     914.20&  1.71[8]&    1.6& $ 6f_{7/2}$&$  5g_{7/2}$&    3167.09&   2.42[7]&   2.9&  $ 5g_{7/2}$&$  7f_{7/2}$&   17639.80&   1.17[5]&    0.5\\
  $ 7p_{1/2}$&$  8d_{3/2}$&     983.14&  2.93[8]&    4.1& $ 6f_{7/2}$&$  5g_{9/2}$&    3169.00&   6.76[8]&   2.9&  $ 5g_{7/2}$&$  7f_{5/2}$&   19362.21&   3.32[6]&    1.2\\
  $ 7p_{1/2}$&$  9s_{1/2}$&     995.13&  2.20[8]&    2.3& $ 8p_{1/2}$&$  8d_{3/2}$&    3644.65&   3.43[8]&   9.4&  $ 8d_{3/2}$&$  7f_{5/2}$&   38451.19&   9.60[5]&    2.0\\
  $ 7p_{3/2}$&$  8d_{5/2}$&    1117.67&  1.81[8]&    5.0& $ 8p_{1/2}$&$  9s_{1/2}$&    3815.10&   1.42[8]&   2.5&  $ 8d_{5/2}$&$  7f_{7/2}$&   39506.95&   9.65[5]&    3.1\\
  $ 7p_{3/2}$&$  8d_{3/2}$&    1124.88&  2.27[7]&    6.0& $ 8p_{3/2}$&$  8d_{5/2}$&    4413.67&   3.18[8]&   0.5&  $ 8d_{5/2}$&$  7f_{5/2}$&   49336.43&   3.23[4]&    2.1\\
\end{tabular}
\end{ruledtabular}
\end{table*}

  We find that
the uncertainties are 0.2-2\% for most of the transitions. Larger uncertainties occur for some of the transitions with
large correlation contributions such as $5f-ng$. Our final results and their uncertainties
 are used to calculate the recommended values of the transition rates,
 oscillator strengths, lifetimes, and the polarizabilities as well as to
evaluate the uncertainties of these results.
\subsection{Transition rates and oscillator strengths}
We combine experimental energies \cite{expt-en} and our final values of the best set matrix elements to calculate
transition rates $A$ and oscillator strengths $f$. The transition rates are calculated using
\begin{equation}
A_{ab}=\frac{2.02613\times10^{18}}{\lambda^3}\frac{S}{2j_a+1}\text{s}^{-1},
\end{equation}
where the wavelength $\lambda$ is in \AA~ and the line strength
$S=d^2$ is in atomic units.

Transition rates $A$ (s$^{-1}$)  for the  60 $ns-np$, $np-nd$, $nd-nf$, and $nf-ng$ transitions
 are given in
Table~\ref{tab-tran}. Vacuum wavelengths obtained from experimental energies \cite{expt-en} are also listed for
reference. The relative uncertainties of the transition rates listed in the column labeled ``Unc.'' are twice the
corresponding matrix element uncertainties since the transition rates are proportional to the squares of the matrix
elements.
 The smallest uncertainties  are for
the $5g-7f$ transitions, while the largest ones  are for the $6d-8p$ and $5f-8d$ transitions owing to large
corresponding uncertainties in the E1 transition matrix elements. We already discussed the importance of the size of
the correlation effects for the dipole matrix element uncertainties. For example, the DF value for the
$5f_{5/2}-7d_{3/2}$ transition (see Table~\ref{tab-dip}) is smaller than the all-order SD value by a factor of 4. The
SDpT value obtained with including partial triple excitations  is larger than the SD value by 4.7\% for this
transition. The scaling procedure decreases both SD and SDpT  values by 4.3\% and 9.6\%, respectively. The
contributions affected by scaling are related to the correlation potential, and therefore, the values of the
correlation energies for the specific state. The scaling coefficients are obtained as a ratio of the ``experimental''
correlation energy (obtained as the difference of the experimental values and the lowest order results) and the
theoretical SD or SDpT correlation energies. Lower accuracy of the theoretical correlation energy leads to larger
scaling effect owing to larger omitted correlation contribution to the matrix elements of the certain class, in
particularly for weaker transition with small DF values.
 \begin{table*}
\caption{\label{tab-osc} Wavelengths $\lambda$ (\AA) and weighted oscillator strengths $gf$ for transitions in Th~IV
calculated using our recommended values of reduced electric-dipole matrix elements and their uncertainties.  The
relative uncertainties  are listed in column ``Unc.'' in \%. In columns ``DF'', we list $f$ values calculated in DF
approximation. In columns ``Expt'', we list  experimental $\lambda$ values \cite{expt-en}.
  In column ``HFR+CP'', we list $f$
values calculated by HFR+CP method \cite{osc-ra}. Numbers in
brackets represent powers of 10.}
\begin{ruledtabular}
\begin{tabular}{llrrclcllllclc}
\multicolumn{2}{c}{Transition} & \multicolumn{1}{c}{$\lambda$} &
\multicolumn{3}{c}{Oscillator Strengths} &
\multicolumn{1}{c}{Unc.} & \multicolumn{2}{c}{Transition} &
\multicolumn{1}{c}{$\lambda$} & \multicolumn{3}{c}{Oscillator
Strengths} &
\multicolumn{1}{c}{Unc.} \\
\multicolumn{1}{c}{Low} & \multicolumn{1}{c}{Upper} &
\multicolumn{1}{c}{Expt} & \multicolumn{1}{c}{DF} &
\multicolumn{1}{c}{HFR+CP} & \multicolumn{1}{c}{Final} &
\multicolumn{1}{c}{(\%)}& \multicolumn{1}{c}{Low} &
\multicolumn{1}{c}{Upper} & \multicolumn{1}{c}{Expt} &
\multicolumn{1}{c}{DF} & \multicolumn{1}{c}{HFR+CP} &
\multicolumn{1}{c}{Final} & \multicolumn{1}{c}{(\%)} \\\hline
   $6d_{5/2}$&$  7f_{7/2}$&     664.13&    1.15[ 0]&   5.4[-1]&    5.27[-1]&    6.7&   $5f_{5/2}$&$  5g_{7/2}$&     627.39&    6.11[-1]&   4.8[-1]&    2.30[-1]&    8.8\\
   $6d_{5/2}$&$  7f_{5/2}$&     666.36&    5.86[-2]&   2.7[-2]&    2.38[-2]&    8.0&   $7p_{3/2}$&$ 10s_{1/2}$&     914.20&    5.35[-2]&   5.6[-2]&    4.30[-2]&    1.6\\
   $6d_{3/2}$&$  6f_{5/2}$&     846.91&    2.57[ 0]&   1.9[ 0]&    1.97[ 0]&    2.0&   $7p_{3/2}$&$  9s_{1/2}$&    1140.61&    1.52[-1]&   1.5[-1]&    1.35[-1]&    3.0\\
   $6d_{5/2}$&$  6f_{7/2}$&     882.39&    3.87[ 0]&   2.6[ 0]&    3.10[ 0]&    1.6&   $7p_{1/2}$&$  7d_{3/2}$&    1682.21&    2.64[ 0]&   2.7[ 0]&    2.15[ 0]&    1.5\\
   $6d_{5/2}$&$  6f_{5/2}$&     886.66&    2.02[-1]&   1.3[-1]&    1.58[-1]&    1.8&   $8p_{1/2}$&$ 10s_{1/2}$&    2086.62&    9.16[-2]&   1.1[-1]&    8.89[-2]&    2.7\\
   $7p_{1/2}$&$  8d_{3/2}$&     983.14&    2.85[-1]&   1.4[-1]&    1.70[-1]&    4.1&   $7d_{5/2}$&$  7f_{7/2}$&    2291.90&    2.69[ 0]&   2.6[ 0]&    2.84[ 0]&    7.2\\
   $6d_{3/2}$&$  7p_{3/2}$&    1565.86&    1.93[-1]&   1.5[-1]&    1.40[-1]&    2.4&   $7d_{5/2}$&$  7f_{5/2}$&    2318.70&    1.41[-1]&   1.3[-1]&    1.54[-1]&    3.0\\
   $6d_{5/2}$&$  7p_{3/2}$&    1707.37&    1.82[ 0]&   1.2[ 0]&    1.35[ 0]&    2.2&   $8p_{3/2}$&$ 10s_{1/2}$&    2349.07&    1.87[-1]&   1.9[-1]&    1.79[-1]&    1.4\\
   $6d_{3/2}$&$  7p_{1/2}$&    1959.02&    1.00[ 0]&   6.0[-1]&    6.98[-1]&    2.8&   $8p_{1/2}$&$  8d_{3/2}$&    3644.65&    2.96[ 0]&   3.5[ 0]&    2.73[ 0]&    9.4\\
   $7s_{1/2}$&$  7p_{3/2}$&    2003.00&    2.42[ 0]&   1.6[ 0]&    1.72[ 0]&    2.6&   $7d_{3/2}$&$  8p_{3/2}$&    4953.85&    2.89[-1]&   3.0[-1]&    2.61[-1]&    0.5\\
   $7p_{3/2}$&$  7d_{5/2}$&    2067.35&    5.20[ 0]&   4.0[ 0]&    4.32[ 0]&    1.4&   $7d_{5/2}$&$  8p_{3/2}$&    5421.88&    2.64[ 0]&   2.5[ 0]&    2.39[ 0]&    0.3\\
   $7p_{3/2}$&$  7d_{3/2}$&    2144.60&    5.84[-1]&   4.3[-1]&    4.82[-1]&    1.4&   $7f_{5/2}$&$  6g_{7/2}$&    5841.02&    1.11[ 1]&   1.0[ 1]&    9.07[ 0]&    5.3\\
   $7p_{3/2}$&$  8s_{1/2}$&    2147.50&    1.34[ 0]&   1.1[ 0]&    1.24[ 0]&    1.6&   $5f_{5/2}$&$  6d_{5/2}$&    6903.05&    1.80[-2]&   1.6[-2]&    7.45[-3]&    7.7\\
   $7d_{3/2}$&$  7f_{5/2}$&    2228.66&    1.68[ 0]&   1.9[ 0]&    1.88[ 0]&    3.4&   $5f_{7/2}$&$  6d_{5/2}$&    9841.54&    2.70[-1]&   2.2[-1]&    1.14[-1]&    7.6\\
   $7s_{1/2}$&$  7p_{1/2}$&    2694.81&    9.48[-1]&   6.0[-1]&    6.60[-1]&    2.8&   $5f_{5/2}$&$  6d_{3/2}$&   10877.55&    1.65[-1]&   1.4[-1]&    6.53[-2]&    8.2\\
   $6f_{5/2}$&$  5g_{7/2}$&    3113.25&    9.64[ 0]&   7.9[ 0]&    7.85[ 0]&    3.0&             &            &           &            &          &            &        \\
   $6f_{7/2}$&$  5g_{7/2}$&    3167.09&    3.55[-1]&   2.9[-1]&    2.91[-1]&    2.9&   $5f_{5/2}$&$  8d_{3/2}$&     617.46&    4.24[-6]&   1.8[-3]&    2.17[-2]&    14\\
   $6f_{7/2}$&$  5g_{9/2}$&    3169.00&    1.24[ 1]&   1.0[ 1]&    1.02[ 1]&    2.9&   $5f_{7/2}$&$  8d_{5/2}$&     632.10&    3.23[-5]&   2.6[-3]&    3.05[-2]&    14\\
   $8p_{3/2}$&$  8d_{5/2}$&    4413.67&    6.00[ 0]&   5.2[ 0]&    5.57[ 0]&    0.5&   $6d_{3/2}$&$  7f_{5/2}$&     643.66&    7.84[-1]&   3.9[-1]&    3.12[-1]&    8.2\\
   $8p_{3/2}$&$  8d_{3/2}$&    4528.35&    6.91[-1]&   5.6[-1]&    6.41[-1]&    0.5&   $5f_{7/2}$&$  5g_{7/2}$&     644.89&    2.49[-2]&   1.7[-2]&    1.03[-2]&    7.8\\
   $8p_{3/2}$&$  9s_{1/2}$&    4794.51&    1.86[ 0]&   1.5[ 0]&    1.76[ 0]&    1.1&   $5f_{7/2}$&$  5g_{9/2}$&     644.97&    8.75[-1]&   6.0[-1]&    3.69[-1]&    7.6\\
   $8s_{1/2}$&$  8p_{3/2}$&    4938.44&    2.82[ 0]&   2.6[ 0]&    2.51[ 0]&    1.8&   $6d_{3/2}$&$  8p_{3/2}$&     765.24&    1.87[-2]&   5.9[-3]&    8.04[-3]&    7.4\\
   $7f_{7/2}$&$  6g_{7/2}$&    6018.30&    4.07[-1]&   3.7[-1]&    3.43[-1]&    3.0&   $6d_{5/2}$&$  8p_{3/2}$&     797.55&    1.62[-1]&   5.0[-2]&    7.00[-2]&    7.3\\
   $7f_{7/2}$&$  6g_{9/2}$&    6018.66&    1.43[ 1]&   1.3[ 1]&    1.20[ 1]&    2.6&   $6d_{3/2}$&$  8p_{1/2}$&     797.94&    6.32[-2]&   2.8[-2]&    1.38[-2]&    16\\
   $8s_{1/2}$&$  8p_{1/2}$&    6713.71&    1.15[ 0]&   9.5[-1]&    1.01[ 0]&    0.9&   $7p_{1/2}$&$ 10s_{1/2}$&     818.32&    2.52[-2]&   3.1[-2]&    2.29[-2]&    0.7\\
   $7d_{3/2}$&$  8p_{1/2}$&    6742.22&    1.35[ 0]&   1.1[ 0]&    1.21[ 0]&    0.4&   $5f_{5/2}$&$  7d_{5/2}$&     823.54&    8.49[-6]&   1.7[-4]&    1.00[-3]&    31\\
   $7p_{3/2}$&$  8d_{5/2}$&    1117.67&    3.75[-1]&   2.3[-1]&    2.03[-1]&    5.0&   $5f_{5/2}$&$  7d_{3/2}$&     835.53&    1.56[-3]&   2.3[-3]&    2.43[-2]&    20\\
   $7d_{3/2}$&$  6f_{5/2}$&   13184.61&    1.41[ 0]&   1.2[ 0]&    1.15[ 0]&    3.0&   $5f_{7/2}$&$  7d_{5/2}$&     853.96&    1.76[-3]&   3.2[-3]&    3.12[-2]&    24\\
   $7d_{5/2}$&$  6f_{7/2}$&   15653.86&    1.71[ 0]&   1.4[ 0]&    1.40[ 0]&    3.0&   $7s_{1/2}$&$  8p_{1/2}$&     897.78&    1.08[-3]&   2.1[-3]&    1.86[-2]&    11\\
   $7d_{5/2}$&$  6f_{5/2}$&   17117.13&    7.72[-2]&   6.5[-2]&    6.28[-2]&    3.1&   $7p_{1/2}$&$  9s_{1/2}$&     995.13&    6.81[-2]&   8.7[-2]&    6.54[-2]&    2.3\\
   $5g_{9/2}$&$  7f_{7/2}$&   17580.87&    1.58[ 0]&   1.5[ 0]&    1.54[ 0]&    1.0&   $7p_{3/2}$&$  8d_{3/2}$&    1124.88&    3.51[-2]&   2.5[-2]&    1.72[-2]&    6.0\\
   $5g_{7/2}$&$  7f_{7/2}$&   17639.80&    4.53[-2]&   4.4[-2]&    4.38[-2]&    0.5&   $7p_{1/2}$&$  8s_{1/2}$&    1684.00&    4.55[-1]&   6.9[-1]&    4.33[-1]&    2.5\\
   $5g_{7/2}$&$  7f_{5/2}$&   19362.21&    1.14[ 0]&   1.1[ 0]&    1.12[ 0]&    1.2&   $8p_{1/2}$&$  9s_{1/2}$&    3815.10&    6.63[-1]&   9.5[-1]&    6.21[-1]&    2.5\\
\end{tabular}
\end{ruledtabular}
\end{table*}
\begin{table}
\caption{\label{tab-life} Lifetimes ${\tau}$
  in Fr-like Th~IV in ns.
 Uncertainties are given in parenthesis.
 The last
column gives relative uncertainties of the final values in \%.
Experimental energies
 \cite{expt-en}
are given in cm$^{-1}$.  The values of lifetimes evaluated in DF approximation are given to illustrate the correlation
contribution. Lifetime of the metastable $7s$ level is given in text.}
\begin{ruledtabular}
\begin{tabular}{llllcl}
 \multicolumn{1}{c}{Level}&
 \multicolumn{1}{c}{Energy}&
 \multicolumn{1}{c}{$\tau^{(\rm DF)}$}&
 \multicolumn{1}{c}{$\tau^{(\rm final)}$}&
 \multicolumn{1}{c}{Unc. (\%)}  &
   \multicolumn{1}{c}{Ref.~\cite{safr-06-th}}\\
\hline
  $6d_{3/2}$&    9193.245  &   431   &        1086(89)   &   8.2&   1090 \\
  $6d_{5/2}$&    14486.34  &   285   &         678(46)    &  6.8&   676  \\
  $7p_{1/2}$&    60239.1
    &   0.764 &        1.099(23)  &   2.1&   1.099\\
  $7p_{3/2}$&    73055.9  &   0.459 &        0.632(10)  &   1.6&   0.632\\
  $8s_{1/2}$&    119621.6 &   0.665 &        0.711(10)  &   1.4&   0.707\\
  $7d_{3/2}$&    119684.6 &   0.563 &        0.665(10)  &   1.5&   0.667\\
  $7d_{5/2}$&    121427.1 &   0.739 &        0.855(14)  &   1.6&   0.854\\
  $6f_{5/2}$&    127269.2  &   0.234 &        0.304(6)   &   1.8&   0.300\\
  $6f_{7/2}$&    127815.3  &   0.241 &        0.301(5)   &   1.6&   0.297\\
  $8p_{1/2}$&    134516.5 &   1.925 &        3.19(14)   &   4.5&   3.194\\
  $8p_{3/2}$&    139870.9 &   1.077 &        1.87(5)    &   2.6&   1.871\\
  $5g_{9/2}$&    159371   &   0.449 &        0.79(3)    &   3.9&   0.780\\
  $5g_{7/2}$&    159390   &   0.455 &        0.83(3)    &   3.9&   0.815\\
  $9s_{1/2}$&    160728.1 &   0.960 &        1.038(13)  &   1.3&   1.031\\
  $8d_{3/2}$&    161954   &   0.980 &        1.19(6)    &   4.9&   1.176\\
  $8d_{5/2}$&    162527.8  &   1.385 &        1.62(5)    &   3.1&   1.600\\
  $7f_{5/2}$&    164554.7  &   0.376 &        0.738(38)  &   5.2&   0.684\\
  $7f_{7/2}$&    165059.0  &   0.384 &        0.689(35)  &   5.1&   0.639\\
  $6g_{9/2}$&    181674    &   0.641 &        1.85(21)   &   11&   1.567\\
  $6g_{7/2}$&    181675   &   0.649 &        2.21(31)   &   14&   1.768\\
\end{tabular}
\end{ruledtabular}
\end{table}

 We present  weighted oscillator strengths
$gf$  calculated using our recommended values of reduced electric-dipole matrix elements $gf^\text{{final}}$ and their
uncertainties in Table~\ref{tab-osc}.  The relative uncertainties  are listed in column ``Unc.'' in \%. In columns
``DF'', we list $gf$ values calculated in DF approximation. In column ``Expt'', we list $\lambda$ recommended by
compilation in Ref.~\cite{expt-en}.  In column ``HFR+CP'', we list $gf$ values calculated by HFR+CP method
\cite{osc-ra}. In the left column of Table~\ref{tab-osc}, we list the 32 transitions when the ``HFR+CP'' values are in
the better agrement with our ``Final'' result.
  Disagreement
between the $gf^\text{{HFR+CP}}$ and $gf^\text{{final}}$
 values is about 2-20\%. The  15 transitions given in the right
column of Table~\ref{tab-osc} are transitions  when the ``HFR+CP'' values are in the better agreement with our $gf$
values obtained in the DF approximation than with our ``Final'' result. The last 16 transitions given in the right
column of Table~\ref{tab-osc} present transitions when the ``HFR+CP'' values disagree with the DF values as well as
with the ``Final'' values by a factor of 2-10 for most of transitions.  It should be noted that among these 16
transitions there are at least 10 transitions with very small  $gf$ values (10$^{-2}$). The uncertainties for such
transitions are significantly larger than the ones for the other transitions shown in Table~\ref{tab-osc}.
\subsection{Lifetimes}
We list the lifetimes of the $(7-9)s$,  $(7-8)p$, $(6-8)d$, $(5-7)f$, and $(5-6)g$  states in Table~\ref{tab-life}.
These values are obtained using the  transition rates listed in Table~\ref{tab-tran}. The uncertainties in the lifetime
values given in parenthesis are obtained from the uncertainties in the matrix elements.
 The column ``Unc.'' gives relative uncertainties of the ¯final values in \%.
 We also list
the lifetime values calculated in DF approximation in Table~\ref{tab-life} to show the size of the correlation
correction for each case.  The energies recommended by Blase and Wyart~\cite{expt-en} are given in column ``Energy''.
Our
 values are compared with the SD calculation of Ref.~\cite{safr-06-th}. The difference with the present results
  are due to more complete inclusion  of the correlation correction in the present work.

  In 2012, the lifetime of the metastable $7s$ level has been measured to be 0.60(7)~s which is in excellent
  agreement with out earlier prediction of 0.59~s ~\cite{safr-06-th}. The
   $7s-6d_{3/2}$ and $7s-6d_{5/2}$ E2 transitions give the only
  significant contributions to the $7s$ lifetime. In this work, we have carried out additional  SDpT and scaled
  calculations of these values and obtained 7.110(47)~a.u. and 9.211(59)~a.u. for the $7s-6d_{3/2}$ and $7s-6d_{5/2}$ E2
  reduced matrix elements, respectively. Our final value of the $7s$ lifetime is 0.570(8)~s.

   \begin{table*}
\caption{\label{tab-scalar} Contributions to the  scalar ($\alpha_0$) and tensor ($\alpha_2$) polarizabilities
 of Th~IV ion in $a_0^3$. Uncertainties are given
in parenthesis. }
\begin{ruledtabular}
\begin{tabular}{lrrlrrlrrrr}
 \multicolumn{3}{c}{${5f_{5/2}}$} &  \multicolumn{3}{c}{${5f_{7/2}}$} &\multicolumn{3}{c}{${6d_{3/2}}$}& \\
 \hline
  \multicolumn{1}{l}{Contr.} &
 \multicolumn{1}{c}{$\alpha_0$}   &
 \multicolumn{1}{c}{$\alpha_2$}   &
\multicolumn{1}{l}{Contr.} &
  \multicolumn{1}{c}{$\alpha_0$}   &
  \multicolumn{1}{c}{$\alpha_2$}   &
     \multicolumn{1}{l}{Contr.} &
\multicolumn{1}{c}{$\alpha_0$}&
 \multicolumn{1}{c}{$\alpha_2$}   &
 \multicolumn{1}{c}{} \\ \hline
  $6d_{3/2}$& 6.21(51)          & -6.21(51) &   $6d_{5/2}$&  6.63(50)  & -6.63(50)   &   $7p_{1/2}$&  3.23(9)  &-3.23(9)  &\\
  $7d_{3/2}$& 0.014(3)          & -0.014(3) &   $7d_{5/2}$&  0.014(3)  & -0.014(3)   &   $8p_{1/2}$&  0.011(2) &-0.011(2) &\\
  $8d_{3/2}$& 0.007(7)          & -0.007(7) &   $8d_{5/2}$&  0.007(7)  & -0.007(7)   &   $9p_{1/2}$&  0.000   & 0.000 &\\
  $nd_{3/2}$& 0.03(3)           & -0.03(3)  &   $nd_{5/2}$&  0.03(3)   & -0.03(3)    &   $np_{1/2}$&  0.02(2) &-0.02(2) &\\
  $6d_{5/2}$& 0.29(2)          &  0.33(3) &   $5g_{7/2}$&  0.003(1)  &  0.003(1)   &   $7p_{3/2}$&  0.413(10) & 0.330(8) &\\
  $7d_{5/2}$& 0.001(1)          &  0.001(1) &   $6g_{7/2}$&  0.001(1)  &  0.001(1)   &   $8p_{3/2}$&  0.006(1) & 0.005(1) &\\
  $nd_{5/2}$& 0.002(2)          &  0.001(1) &   $ng_{7/2}$&  0.006(6)  &  0.007(7)   &   $np_{3/2}$&  0.008(8) & 0.006(6) &\\
  $5g_{7/2}$& 0.07(1)         & -0.026(2)    &   $5g_{9/2}$&  0.09(1) & -0.043(3)   &   $5f_{5/2}$& -9.31(76) & 1.86(5)  &\\
  $6g_{7/2}$& 0.02(2)         & -0.005(1)    &   $6g_{9/2}$&  0.03(1) & -0.012(2)   &   $6f_{5/2}$&  1.70(3)  &-0.341(7) &\\
  $7g_{7/2}$& 0.02(1)         & -0.007(1)    &   $7g_{9/2}$&  0.03(1) & -0.012(1)   &   $7f_{5/2}$&  0.156(13) &-0.031(3) &\\
  $ng_{7/2}$& 0.13(13)          & -0.05(5) &   $ng_{9/2}$ &  0.10(10)  & -0.05(5)    &   $nf_{5/2}$&  0.22(11) &-0.19(8) &\\
  Tail      &   0.15(20)        &  -0.05(7) &   Tail      &   0.17(20) &  -0.07(9)   &   Tail      &   0.02(5)   & -0.005(11)   &\\
  Core      &  7.75(7)          &           &   Core      &   7.75(7)  &             &    Core      &   7.75(7)    &\\
  vc        &   -0.02(1)        &           &   vc        &  -0.02(1)  &             &    vc        &  -0.43(7)   &\\
  Total     & 14.67(60)         & -6.07(53) &   Total     &  14.84(59) &  -6.95(52)  &   Total     &   4.53(81)&-1.62(21)
  &\\
\hline
 \multicolumn{3}{c}{${6d_{5/2}}$} & \multicolumn{3}{c}{${7p_{3/2}}$} &
\multicolumn{2}{c}{${7p_{1/2}}$} & \multicolumn{2}{c}{${7s_{1/2}}$}\\
 \multicolumn{1}{l}{Contr.} &
 \multicolumn{1}{c}{$\alpha_0$}   &
  \multicolumn{1}{c}{$\alpha_2$}   &
 \multicolumn{1}{l}{Contr.} &
 \multicolumn{1}{c}{$\alpha_0$}   &
 \multicolumn{1}{c}{$\alpha_2$}   &
\multicolumn{1}{l}{Contr.} &
 \multicolumn{1}{c}{$\alpha_0$}   &
 \multicolumn{1}{r}{Contr.} &
 \multicolumn{1}{c}{$\alpha_0$} \\
 \hline
$7p_{3/2}$&  3.16(7)   & -3.16(7)  & $7s_{1/2}$&  -8.31(22)  &  8.31(22)  & $7s_{1/2}$&-11.54(32)&  $7p_{1/2}$&  11.54(32)   \\
$8p_{3/2}$&  0.036(3)  & -0.036(3) & $8s_{1/2}$&   6.90(11)  & -6.90(11)  & $8s_{1/2}$&  2.96(7) &  $8p_{1/2}$&   0.036(4)   \\
$9p_{3/2}$&  0.003     & -0.003    & $9s_{1/2}$&   0.212(6)  & -0.212(6)  & $9s_{1/2}$&  0.16    &  $np_{1/2}$&   0.08(8)   \\
$np_{3/2}$&  0.05(5)   & -0.05(5)  & $ns_{1/2}$&   0.10(1)   & -0.10(1)   & $ns_{1/2}$&  0.09(2) &  $7p_{3/2}$&  16.62(43)   \\
$5f_{5/2}$& -0.29(2) & -0.33(3)& $6d_{3/2}$&  -0.413(10) & -0.330(8)  & $6d_{3/2}$& -6.45(18) &  $8p_{3/2}$&   0.003(2)   \\
$6f_{5/2}$&  0.100(2)  &  0.114(2) & $7d_{3/2}$&   2.67(4)   &  2.14(3) & $7d_{3/2}$& 14.64(22)&  $np_{3/2}$&   0.05(5)   \\
$nf_{5/2}$&  0.05(3)   & -0.06(3)  & $nd_{3/2}$&   0.03(2)   &  0.03(2)   & $8d_{3/2}$&  0.40(7) &          &     \\
$5f_{7/2}$& -8.84(67)  &  3.16(24) & $6d_{5/2}$&  -4.74(11)  &  0.95(2) &  $nd_{3/2}$&  0.12(6)&          &      \\
$6f_{7/2}$&   1.94(3)  & -0.69(1)& $7d_{5/2}$&  22.21(30)    & -4.44(6)   &           &          &            &             \\
$7f_{7/2}$&  0.19(1) & -0.067(4) & $8d_{5/2}$&  0.31(2)  & -0.06(1) &           &          &            &             \\
$nf_{7/2}$&  0.21(11) & -0.08(4) &$nd_{5/2}$&   0.07(7)   & -0.01(1)   &             &            &           &           \\
Tail      &   0.02(4) & -0.01(1)& Tail      & 0.01(2)   & -0.008(6) & Tail      & 0.01(3)    &  Tail      &   0.003(1)  \\
Core      &  7.75(7)  &           & Core      & 7.75(7)     &           &  Core      &  7.75(7)   &    Core      &  7.75(7)   \\
vc        &   -0.72(4)&           & vc        & 0.001       &           &   vc       & -0.004(1)  &     vc       &  -0.50(10) \\
Total     &  3.67(70) & -1.09(27)  & Total     & 26.88(42)   & -0.64(26)  & Total     &8.13(41)  &  Total     &35.58(55) \\
\end{tabular}
\end{ruledtabular}
\end{table*}
\section{ Scalar and tensor excited state polarizabilities}
The valence scalar $\alpha_{0}(v)$  and tensor $\alpha_{2}(v)$ polarizabilities of an excited state $v$ of Th~IV are
given by
\begin{equation}
\alpha _{\text{0}}(v)\ =\frac{2}{3(2j_{v}+1)}\sum_{nlj}\frac{|\langle v||rC_{1}||nlj\rangle
|^{2}}{E_{nlj}-E_{v}},\label{eqp1}
\end{equation}
\begin{align}
\alpha _{2}(v)& \ =(-1)^{j_{v}}\sqrt{\frac{40j_{v}(2j_{v}-1)}{%
3(j_{v}+1)(2j_{v}+1)(2j_{v}+3)}}\   \label{eq1a} \nonumber\\
& \times \sum_{nlj}(-1)^{j}\left\{
\begin{array}{lll}
j_{v} & 1 & j \\
1 & j_{v} & 2
\end{array}
\right\} \frac{|\langle v||rC_{1}||nlj\rangle
|^{2}}{E_{nlj}-E_{v}}\ .
\end{align}
where  $C_{1}(\hat{r})$ is a normalized spherical harmonic and the sum over $nlj$ runs over all states with allowed
electric-dipole transitions to a state $v$ \cite{multipol}. The reduced matrix elements in the dominant contributions
to the above sum are evaluated using out final values of the dipole matrix elements and available experimental energies
\cite{expt-en}. The uncertainties in the polarizability contributions are obtained from the uncertainties in the matrix
elements. We use theoretical SD energies and SD wave functions to evaluate terms with $n<26$ in Eqs.~(\ref{eqp1}) and
(\ref{eq1a}). The remaining contributions to $\alpha_{0}$ and $\alpha_{2}$ from orbitals with $27 \leq n \leq 70$ are
evaluated in the RPA approximation since the contributions from these terms are smaller than 0.01\% in all cases. These
terms are grouped together as ``Tail''. Their uncertainty is estimated as the difference of the corresponding DF and
RPA values. We note that the $ng$ states with $n>18$, $nf$ states with $n>19$, $nd$ states with $n>20$, and $np$, $ns$
states with $n>21$  have positive energies and provide a discrete representation of the continuum in our basis.

 We list the contributions to the scalar polarizabilities of
the $5f_j$, $6d_j$, $7p_j$, and $7s$  states  and tensor polarizabilities of the
 $5f_j$, $6d_j$, and $7p_{3/2}$ states
 in Table~\ref{tab-scalar}. The dominant contributions are listed
separately. The remaining contributions are grouped together. For example, ``$nd_{3/2}$'' contribution includes all of
the $nd_{3/2}$ terms with $n\leq 26$ excluding only the terms that were already listed separately.

We evaluate the contribution from the ionic core
  $\alpha _{\text{core}}$
in the RPA and find $\alpha _{\textrm{core}}= 7.75(7)$~a.u. We estimate uncertainty in this term to be on the order of
1\% based on the comparison of the RPA values for heavier noble gases (Kr and Xe) with experiment and comparison of the
ionic core RPA values for  heavy ions (such as Ba$^{2+}$) with  coupled-cluster results (see Table 4 of
Ref.~\cite{MitSafCla10}). Our result is in excellent agreement with the recent RESIS measurement of Th$^{4+}$
polarizability of 7.702(6)~a.u. \cite{18192EL}. A counter term $\alpha _{\rm vc}$ compensating for excitation from the
core to the valence shell which violates the Pauli principle is also evaluated in the RPA and is given in rows labelled
 ``vc'' in Table~\ref{tab-scalar}. A difference of the DF and RPA values is taken to be its uncertainly. The core
polarizability gives a very large contribution to all scalar polarizabilities, ranging from nearly 100\% for the
7p$_{1/2}$, where valence terms cancel out each other, to 20\% for the $7s$ state. Its contribution to the ground state
$5f_{5/2}$ polarizability is 53\%. For comparison, the core polarizability contributes only 6\% to the total $7s$
ground state polarizability of Fr.

The evaluations of the $\alpha_{0}$ and $\alpha_{2}$ polarizabilities differ only in the angular part. Both scalar and
tensor ground state valence polarizabilities are dominated by a single transition, $5f_{5/2}-6d_{3/2}$. It contributes
89\% of the scalar valence polarizability. Its contribution (-6.21~a.u.) to the tensor polarizability is larger than
the total value, since  the $5f_{5/2}-6d_{5/2}$ contributes 0.33~a.u. with the opposite sign. The continuous part of
spectra is responsible for 1\% of $\alpha_{0}$ and $\alpha_{2}$ for the $5f_{5/2}$ state.  We discuss comparison of the
ground state polarizability values with RESIS experiments~\cite{lundeen} in Section~\ref{resis}.

 The dominant
contribution, 98.5\%, to the $\alpha_{2}(5f_{7/2})$ value comes from the  $nd_{5/2}$ states, particulary from the
$6d_{5/2}$ state.  The contributions to the $\alpha_{2}(5f_{7/2})$  value from the  $ng_{7/2}$ and $ng_{9/2}$ states
are 0.2\% and 1.7\%, respectively, and have a different sign. The dominant contributions to the $\alpha_{2}(6d_{3/2})$
value are from the  $7p_{1/2}$ and
 $5f_{5/2}$ states and they partly cancel each other. All other
 states contribute to the  $\alpha_{2}(6d_{3/2})$  value only
 15\%. The dominant contributions
to the $\alpha_{2}(6d_{5/2})$  value are  from the  $7p_{3/2}$ and
 $5f_{7/2}$ states and they nearly exactly cancel each other (to 0.1\%).
 As a result,  the  $\alpha_{2}(6d_{5/2})$ polarizability value mainly comes from the  $5f_{5/2}$ and  $6f_{7/2}$
states. The contributions to the $\alpha_{2}(7p_{3/2})$  value are distributed between the $7s_{1/2}$, $8s$,
$7d_{3/2}$, $6d_{5/2}$, and $7d_{5/2}$ states. There are very large  cancelations among these five terms.

 The uncertainties in the values of the $5f$ and $6d$ polarizabilities are overwhelmingly dominated by the uncertainties in the
 $5f-6d$ transitions. Accurate measurement of the $6d$ lifetimes would allow to significantly reduce all of the uncertainties.

\section{Quadrupole moment}
The electric quadrupole moment $\Theta(\gamma J)$ of an atom in electronic state $| \gamma J \rangle$ is defined as the
diagonal matrix element of the $q=0$ component of the electric quadrupole operator $Q$ in a spherical basis
\begin{equation}
\Theta(\gamma J)= \left<\Psi( \gamma JM_J) \left|  Q_0 \right| \Psi( \gamma J M_J)\right>, \label{eqq}
\end{equation}
with the magnetic quantum number $M_J$ taken to be equal to its maximum value, $M_J=J$ \cite{I06}. The quadrupole
moment is expressed via the reduced matrix element of the quadrupole operator as
\begin{equation}
\Theta(\gamma J)=\frac{(2J)!}{\sqrt{(2J-2)!(2J+3)!}}
 \left<\Psi( \gamma J) \left\|  Q \right\|
 \Psi( \gamma J )\right>.
\end{equation}
Therefore, the calculation of the quadrupole moment of the ground state of Th~IV reduces to the calculation of the
diagonal matrix element of the electric-quadrupole operator.

 The summary of the calculations is given in
Table~\ref{quad}, where we list the results of the lowest-order DF, third-order many-body perturbation theory MBPT3,
and all-order SD, SDpT, and SD$_{sc}$ calculations. The MBPT3(6) and MBPT3(10) columns give the third-order values
calculated with $l_{max}=6$ and $l_{max}=10$, respectively. The difference of these values gives an estimate of the
higher-partial wave  contributions, and is added to the all-order values which were obtained with $l_{max}=6$. It
contributes 1\% lowering the quadrupole moment value. We can not use the difference of the \textit{ab initio} and
scaled all-order results to accurately estimate the uncertainty of the final value, since the  correction terms
affected by the scaling account for only 1/3 of the total correlation. Instead, we use the calculation of the
quadrupole moments in $nd$ states of Ca$^+$, Sr$^+$, and Ba$^+$ \cite{JiaAroSaf08}, where the uncertainties could be
accurately estimated. The Ca$^+$  theoretical ground state quadrupole moment \cite{JiaAroSaf08} is in excellent
agreement with a precision experiment ~\cite{quad2}. We find that the uncertainties of theoretical  values reported in
\cite{JiaAroSaf08} were about 3\% of the correlation correction for Ca$^+$ and 2.3\% for Ba$^+$, while the correlation
corrections contributed 25\% and 17\% for these ions. In the present Th~IV case, the correlation contributed 35\% to
the ground state quadrupole moment, so we estimate that it is accurate to about 4\%, yielding the final value of
0.624(14)~a.u.

\begin{table}
 \caption{\label{quad} Quadrupole moment $\Theta$ of Th~IV in the ground $5f_{5/2}$ state in a.u.. See text for
designations.}
\begin{ruledtabular}
\begin{tabular}{cccccc}
\multicolumn{1}{c}{DF} & \multicolumn{1}{c}{MBPT3(6)} &\multicolumn{1}{c}{MBPT3(10)} & \multicolumn{1}{c}{SD} &
\multicolumn{1}{c}{SDpT} &\multicolumn{1}{c}{SD$_{sc}$}  \\
 \hline
0.916 &  0.555  & 0.550   &0.620&   0.628&   0.624(14)\\
\end{tabular}
\end{ruledtabular}
\end{table}

\section{Comparison with RESIS values}
\label{resis}
\begin{table}
 \caption{\label{comp} Comparison of the electric-dipole $5f_{5/2}-6d_j$ matrix elements $D$, ground state
 quadrupole moment $\Theta$, scalar $\alpha_0$ and tensor $\alpha_2$ ground state Th$^{3+}$ polarizabilities, and
 Th$^{4+}$ ground state polarizability with the RESIS experimental results \cite{18192EL,lundeen}.   $R$ is the ratio of
 the $5f_{5/2}-6d_{3/2}$ and  $5f_{5/2}-6d_{5/2}$ matrix elements.   $\alpha^{\textrm{mod}}$ is the polarizability with
 the contribution of the $5f_{5/2}-6d$ terms subtracted out.  $\alpha^{\textrm{mod1}}_0$ has core polarizability
 $\alpha_{\textrm{core}}$ subtracted
 out as well.   All values are in atomic units. }
\begin{ruledtabular}
\begin{tabular}{lrr}
\multicolumn{1}{c}{Property} & \multicolumn{1}{c}{Present} &\multicolumn{1}{c}{Refs.~\cite{18192EL,lundeen}}  \\
\hline
$\Theta$                                      & 0.624(14) &  0.54(4) \\
$|\langle 5f_{5/2} ||D|| 6d_{3/2}\rangle|$    & 1.530(63) &  1.435(10) \\
$|\langle 5f_{5/2} ||D|| 6d_{5/2}\rangle|$    & 0.412(16) &  0.414(24) \\
$R$                                           & 3.716(23) &  3.47(20)\\
$\alpha_0$                                    & 14.67(60) &   15.42(17) \\
$\alpha^{\textrm{mod}}_0$                     & 8.18(34)  &   9.67(15) \\
$\alpha_{\textrm{core}}$ [Th$^{4+}$]          & 7.75(7)   &   7.702(6)\\
$\alpha^{\textrm{mod1}}_0$                    & 0.43(33)  &   1.97(15)  \\
$\alpha_2$                                    & -6.07(53) &   -3.6(1.3) \\
$\alpha^{\textrm{mod}}_2$                     &  -0.19(13)&   1.5(1.3) \\
\end{tabular}
\end{ruledtabular}
\end{table}
Binding energies of high-L Rydberg states ($L\geq7$) of Th$^{2+}$ with $n$ = 27-29 were studied using the resonant
excitation Stark ionization spectroscopy (RESIS) method in \cite{lundeen}.
    Analysis of the observed RESIS spectra led to determination
of five properties of the Th$^{3+}$ ion: its electric quadrupole moment $\Theta$ in the ground state, adiabatic scalar
and tensor ground state dipole polarizabilities, and the dipole matrix elements connecting the ground $5f_{5/2}$ level
to the low-lying $6d_{3/2}$ and $6d_{5/2}$ levels. The frequencies of the 14 well-resolved single lines were fit to
determine the best values of the following parameters \cite{lundeen}: $\langle 5f_{5/2} ||D|| 6d_{j}\rangle$ matrix
elements, $\Theta$, and scalar and tensor polarizabilities $\alpha^{\textrm{mod}}_{0,2}$ with
 the contribution of the $5f_{5/2}-6d$ terms subtracted out. The results of the fit were used to determine full adiabatic
 polarizabilities $\alpha_0$ and $\alpha_2$. The core polarizability $\alpha_{\textrm{core}}$, i.e. the polarizability
 of Th$^{4+}$ was determined in Ref.~\cite{18192EL}. We list  comparison of our results with RESIS data for all of
 these quantities in Table~\ref{comp}. We have already discussed the calculations of these properties and their
 uncertainties in the  previous section, so we discuss only comparison of the results here.
 The present values and the RESIS fit results for the
 quadrupole moment
$\Theta$ and $\langle 5f_{5/2} ||D|| 6d_{3/2}\rangle$  agree to 2$\sigma$ and 1.5$\sigma$, respectively. The central
values for the $\langle 5f_{5/2} ||D|| 6d_{5/2}\rangle$ are nearly identical. This leads to difference in ratio $R$ of
the $5f_{5/2}-6d_{3/2}$ and $5f_{5/2}-6d_{5/2}$ matrix elements. The theoretical prediction for this ratio is by far
more accurate (0.6\%) than the theory values of the matrix elements (4\%), since the correlation corrections are very
similar for the transitions involving states of the same fine-structure multiplet. We took the difference of the ratios
calculated using third order MBPT and all-order methods as the uncertainty, which is rather conservative. Using our
value of the ratio and RESIS $5f_{5/2}-6d_{3/2}$ matrix element yields 0.386(4)~a.u. for the $5f_{5/2}-6d_{5/2}$ matrix
element, which is shifted by 1$\sigma$ from the RESIS fit  value of 0.414(24)~a.u.

The $\alpha^{\textrm{mod}}_0$ value is dominated by the core polarizability, therefore, we separated it out for
comparison purposes:
$$
\alpha^{\textrm{mod}}_0=\alpha_{\textrm{core}}+\alpha^{\textrm{mod1}}_0.
$$
The theoretical value for the core polarizability is in excellent agreement with the experimental value \cite{18192EL}.
The remainder $\alpha^{\textrm{mod1}}_0$ disagrees significantly with the RESIS fit. Large fraction (65\%) of this
remainder
 contribution and essentially all of its uncertainty comes from the $ng_{7/2}$ terms with $n>7$. Even if we add  DF
value for $(n>7)g_{7/2}$ terms, 0.70~a.u. (which is an upper bound for this property since DF systematically and
significantly overestimates the polarizability contributions),  to the remaining contributions, we get
$\alpha^{\textrm{mod1}}_0=0.86$~a.u.
 Therefore it is difficult to come up with a scenario in which
$\alpha^{\textrm{mod1}}_0$ is as high as 1.97~a.u. The total theoretical $\alpha_0$ and   $\alpha_2$ values are in
agreement with RESIS values to about  2$\sigma$. It would be very interesting to see if RESIS line data can be
reproduced by using only one free parameter, $\langle 5f_{5/2} ||D|| 6d_{3/2}\rangle$, and allowing
$\Theta=0.624(14)$~a.u., $R=3.716(23)$~a.u., $\alpha_{\textrm{core}}=7.702(6)$~a.u.,
$\alpha^{\textrm{mod1}}_0=0.43(33)$~a.u., $\alpha^{\textrm{mod}}_2= -0.19(12)$~a.u. to vary within the 1-2$\sigma$
uncertainties.
\section{Conclusion}
In summary, we carried out a  systematic study of Fr-like Th~IV atomic
properties for the $7s$, $8s$, $9s$, $10s$, $7p$, $8p$, $6d$, $7d$, $5f$, $6f$, $7f$, $5g$, and $6g$
 states using
high-precision relativistic all-order approach.
  Reduced matrix
elements, oscillator strengths, transition rates, and lifetimes for the 24 first low-lying levels, ground state
quadrupole moment, scalar polarizabilities of the seven first states, and tensor polarizabilities of the $5f$, $6d$,
$7p_{3/2}$ states are calculated.  We evaluate the uncertainties of our calculations for all of the values listed in
this work. Detailed comparison of the present values with RESIS experimental results ~\cite{18192EL,lundeen} is carried
out. These calculations provide recommended values critically evaluated for their accuracy for the development of ultra
precise nuclear clock, RESIS experiments with actinide ions, and other studies.

\begin{acknowledgments}
 The work of M.S.S.
was supported in part by National Science Foundation Grant No.\ PHY-1068699.
\end{acknowledgments}

%\bibliography{thIV}

\begin{thebibliography}{30}
\expandafter\ifx\csname natexlab\endcsname\relax\def\natexlab#1{#1}\fi \expandafter\ifx\csname
bibnamefont\endcsname\relax
  \def\bibnamefont#1{#1}\fi
\expandafter\ifx\csname bibfnamefont\endcsname\relax
  \def\bibfnamefont#1{#1}\fi
\expandafter\ifx\csname citenamefont\endcsname\relax
  \def\citenamefont#1{#1}\fi
\expandafter\ifx\csname url\endcsname\relax
  \def\url#1{\texttt{#1}}\fi
\expandafter\ifx\csname urlprefix\endcsname\relax\def\urlprefix{URL }\fi \providecommand{\bibinfo}[2]{#2}
\providecommand{\eprint}[2][]{\url{#2}}

\bibitem[{\citenamefont{Sonnenschein et~al.}(2012)\citenamefont{Sonnenschein,
  Raeder, Hakimi, Moore, and Wendt}}]{1}
\bibinfo{author}{\bibfnamefont{V.}~\bibnamefont{Sonnenschein}},
  \bibinfo{author}{\bibfnamefont{S.}~\bibnamefont{Raeder}},
  \bibinfo{author}{\bibfnamefont{A.}~\bibnamefont{Hakimi}},
  \bibinfo{author}{\bibfnamefont{I.~D.} \bibnamefont{Moore}}, \bibnamefont{and}
  \bibinfo{author}{\bibfnamefont{K.}~\bibnamefont{Wendt}}, \bibinfo{journal}{J.
  Phys. B} \textbf{\bibinfo{volume}{45}}, \bibinfo{pages}{165005}
  (\bibinfo{year}{2012}).

\bibitem[{\citenamefont{Raeder et~al.}(2011)\citenamefont{Raeder, Sonnenschein,
  Gottwald, Moore, Reponen, Rothe, Trautmann, and Wendt}}]{2}
\bibinfo{author}{\bibfnamefont{S.}~\bibnamefont{Raeder}},
  \bibinfo{author}{\bibfnamefont{V.}~\bibnamefont{Sonnenschein}},
  \bibinfo{author}{\bibfnamefont{T.}~\bibnamefont{Gottwald}},
  \bibinfo{author}{\bibfnamefont{I.~D.} \bibnamefont{Moore}},
  \bibinfo{author}{\bibfnamefont{M.}~\bibnamefont{Reponen}},
  \bibinfo{author}{\bibfnamefont{S.}~\bibnamefont{Rothe}},
  \bibinfo{author}{\bibfnamefont{N.}~\bibnamefont{Trautmann}},
  \bibnamefont{and} \bibinfo{author}{\bibfnamefont{K.}~\bibnamefont{Wendt}},
  \bibinfo{journal}{J. Phys. B} \textbf{\bibinfo{volume}{44}},
  \bibinfo{pages}{165005} (\bibinfo{year}{2011}).

\bibitem[{\citenamefont{Porsev et~al.}(2010)\citenamefont{Porsev, Flambaum,
  Peik, and {Chr. Tamm}}}]{peik}
\bibinfo{author}{\bibfnamefont{S.~G.} \bibnamefont{Porsev}},
  \bibinfo{author}{\bibfnamefont{V.}~\bibnamefont{Flambaum}},
  \bibinfo{author}{\bibfnamefont{E.}~\bibnamefont{Peik}}, \bibnamefont{and}
  \bibinfo{author}{\bibnamefont{{Chr. Tamm}}}, \bibinfo{journal}{Phys. Rev.
  Lett.} \textbf{\bibinfo{volume}{105}}, \bibinfo{pages}{182501}
  (\bibinfo{year}{2010}).

\bibitem[{\citenamefont{Campbell et~al.}(2009)\citenamefont{Campbell, Steele,
  Churchill, DePalatis, Naylor, Matsukevich, Kuzmich, and Chapman}}]{gt2}
\bibinfo{author}{\bibfnamefont{C.~J.} \bibnamefont{Campbell}},
  \bibinfo{author}{\bibfnamefont{A.~V.} \bibnamefont{Steele}},
  \bibinfo{author}{\bibfnamefont{L.~R.} \bibnamefont{Churchill}},
  \bibinfo{author}{\bibfnamefont{M.~V.} \bibnamefont{DePalatis}},
  \bibinfo{author}{\bibfnamefont{D.~E.} \bibnamefont{Naylor}},
  \bibinfo{author}{\bibfnamefont{D.~N.} \bibnamefont{Matsukevich}},
  \bibinfo{author}{\bibfnamefont{A.}~\bibnamefont{Kuzmich}}, \bibnamefont{and}
  \bibinfo{author}{\bibfnamefont{M.~S.} \bibnamefont{Chapman}},
  \bibinfo{journal}{Phys. Rev. Lett.} \textbf{\bibinfo{volume}{102}},
  \bibinfo{pages}{233004} (\bibinfo{year}{2009}).

\bibitem[{\citenamefont{{Campbell} et~al.}(2012)\citenamefont{{Campbell},
  {Radnaev}, {Kuzmich}, {Dzuba}, {Flambaum}, and {Derevianko}}}]{CamRadKuz12}
\bibinfo{author}{\bibfnamefont{C.~J.} \bibnamefont{{Campbell}}},
  \bibinfo{author}{\bibfnamefont{A.~G.} \bibnamefont{{Radnaev}}},
  \bibinfo{author}{\bibfnamefont{A.}~\bibnamefont{{Kuzmich}}},
  \bibinfo{author}{\bibfnamefont{V.~A.} \bibnamefont{{Dzuba}}},
  \bibinfo{author}{\bibfnamefont{V.~V.} \bibnamefont{{Flambaum}}},
  \bibnamefont{and}
  \bibinfo{author}{\bibfnamefont{A.}~\bibnamefont{{Derevianko}}},
  \bibinfo{journal}{Physical Review Letters} \textbf{\bibinfo{volume}{108}},
  \bibinfo{eid}{120802} (\bibinfo{year}{2012}).

\bibitem[{\citenamefont{Berengut et~al.}(2009)\citenamefont{Berengut, Dzuba,
  Flambaum, and Porsev}}]{5}
\bibinfo{author}{\bibfnamefont{J.~C.} \bibnamefont{Berengut}},
  \bibinfo{author}{\bibfnamefont{V.~A.} \bibnamefont{Dzuba}},
  \bibinfo{author}{\bibfnamefont{V.~V.} \bibnamefont{Flambaum}},
  \bibnamefont{and} \bibinfo{author}{\bibfnamefont{S.~G.}
  \bibnamefont{Porsev}}, \bibinfo{journal}{Phys. Rev. Lett.}
  \textbf{\bibinfo{volume}{102}}, \bibinfo{pages}{210801}
  (\bibinfo{year}{2009}).

\bibitem[{\citenamefont{Campbell et~al.}(2011)\citenamefont{Campbell, Radnaev,
  and Kuzmich}}]{gt3}
\bibinfo{author}{\bibfnamefont{C.~J.} \bibnamefont{Campbell}},
  \bibinfo{author}{\bibfnamefont{A.~G.} \bibnamefont{Radnaev}},
  \bibnamefont{and} \bibinfo{author}{\bibfnamefont{A.}~\bibnamefont{Kuzmich}},
  \bibinfo{journal}{Phys. Rev. Lett.} \textbf{\bibinfo{volume}{106}},
  \bibinfo{pages}{223001} (\bibinfo{year}{2011}).

\bibitem[{\citenamefont{{Porsev} and {Flambaum}}(2010)}]{eb}
\bibinfo{author}{\bibfnamefont{S.~G.} \bibnamefont{{Porsev}}} \bibnamefont{and}
  \bibinfo{author}{\bibfnamefont{V.~V.} \bibnamefont{{Flambaum}}},
  \bibinfo{journal}{Phys. Rev. A} \textbf{\bibinfo{volume}{81}},
  \bibinfo{eid}{032504} (\bibinfo{year}{2010}).

\bibitem[{\citenamefont{Gerstenkorn et~al.}(1974)\citenamefont{Gerstenkorn,
  Luc, Verges, Englekemeir, Gindler, and Tomkins}}]{nuc}
\bibinfo{author}{\bibfnamefont{S.}~\bibnamefont{Gerstenkorn}},
  \bibinfo{author}{\bibfnamefont{P.}~\bibnamefont{Luc}},
  \bibinfo{author}{\bibfnamefont{J.}~\bibnamefont{Verges}},
  \bibinfo{author}{\bibfnamefont{D.~W.} \bibnamefont{Englekemeir}},
  \bibinfo{author}{\bibfnamefont{J.~E.} \bibnamefont{Gindler}},
  \bibnamefont{and} \bibinfo{author}{\bibfnamefont{F.~S.}
  \bibnamefont{Tomkins}}, \bibinfo{journal}{J. Phys. (Paris)}
  \textbf{\bibinfo{volume}{35}}, \bibinfo{pages}{483} (\bibinfo{year}{1974}).

\bibitem[{\citenamefont{{Radnaev} et~al.}(2012)\citenamefont{{Radnaev},
  {Campbell}, and {Kuzmich}}}]{RadCamKuz12}
\bibinfo{author}{\bibfnamefont{A.~G.} \bibnamefont{{Radnaev}}},
  \bibinfo{author}{\bibfnamefont{C.~J.} \bibnamefont{{Campbell}}},
  \bibnamefont{and}
  \bibinfo{author}{\bibfnamefont{A.}~\bibnamefont{{Kuzmich}}},
  \bibinfo{journal}{\pra} \textbf{\bibinfo{volume}{86}}, \bibinfo{eid}{060501}
  (\bibinfo{year}{2012}).

\bibitem[{\citenamefont{Keele et~al.}(2011)\citenamefont{Keele, Hanni, Woods,
  Lundeen, and Fehrenbach}}]{lundeen}
\bibinfo{author}{\bibfnamefont{J.~A.} \bibnamefont{Keele}},
  \bibinfo{author}{\bibfnamefont{M.~E.} \bibnamefont{Hanni}},
  \bibinfo{author}{\bibfnamefont{S.~L.} \bibnamefont{Woods}},
  \bibinfo{author}{\bibfnamefont{S.~R.} \bibnamefont{Lundeen}},
  \bibnamefont{and} \bibinfo{author}{\bibfnamefont{C.~W.}
  \bibnamefont{Fehrenbach}}, \bibinfo{journal}{Phys. Rev. A}
  \textbf{\bibinfo{volume}{83}}, \bibinfo{pages}{062501}
  (\bibinfo{year}{2011}).

\bibitem[{\citenamefont{Blaise and Wyart}(1992)}]{expt-en}
\bibinfo{author}{\bibfnamefont{J.}~\bibnamefont{Blaise}} \bibnamefont{and}
  \bibinfo{author}{\bibfnamefont{J.-F.} \bibnamefont{Wyart}}, in
  \emph{\bibinfo{booktitle}{International Tables of Selected Constants, Vol.
  20}} (\bibinfo{publisher}{Centre National de la Recherche Scientifique},
  \bibinfo{address}{Paris}, \bibinfo{year}{1992}).

\bibitem[{\citenamefont{Safronova et~al.}(2007)\citenamefont{Safronova,
  Johnson, and Safronova}}]{safr-07-fr}
\bibinfo{author}{\bibfnamefont{U.~I.} \bibnamefont{Safronova}},
  \bibinfo{author}{\bibfnamefont{W.~R.} \bibnamefont{Johnson}},
  \bibnamefont{and} \bibinfo{author}{\bibfnamefont{M.~S.}
  \bibnamefont{Safronova}}, \bibinfo{journal}{Phys.\ Rev.\ A}
  \textbf{\bibinfo{volume}{76}}, \bibinfo{pages}{042504}
  (\bibinfo{year}{2007}).

\bibitem[{\citenamefont{Safronova et~al.}(2006)\citenamefont{Safronova,
  Johnson, and Safronova}}]{safr-06-th}
\bibinfo{author}{\bibfnamefont{U.~I.} \bibnamefont{Safronova}},
  \bibinfo{author}{\bibfnamefont{W.~R.} \bibnamefont{Johnson}},
  \bibnamefont{and} \bibinfo{author}{\bibfnamefont{M.~S.}
  \bibnamefont{Safronova}}, \bibinfo{journal}{Phys.\ Rev.\ A}
  \textbf{\bibinfo{volume}{74}}, \bibinfo{pages}{042511}
  (\bibinfo{year}{2006}).

\bibitem[{\citenamefont{Migdalek and Glowacz-Proszkiewicz}(2007)}]{migd-07}
\bibinfo{author}{\bibfnamefont{J.}~\bibnamefont{Migdalek}} \bibnamefont{and}
  \bibinfo{author}{\bibfnamefont{A.}~\bibnamefont{Glowacz-Proszkiewicz}},
  \bibinfo{journal}{J. Phys. B} \textbf{\bibinfo{volume}{40}},
  \bibinfo{pages}{4143} (\bibinfo{year}{2007}).

\bibitem[{\citenamefont{Bi\'{e}mont et~al.}(2004)\citenamefont{Bi\'{e}mont,
  Fivet, and Quinet}}]{osc-ra}
\bibinfo{author}{\bibfnamefont{E.}~\bibnamefont{Bi\'{e}mont}},
  \bibinfo{author}{\bibfnamefont{V.}~\bibnamefont{Fivet}}, \bibnamefont{and}
  \bibinfo{author}{\bibfnamefont{P.}~\bibnamefont{Quinet}},
  \bibinfo{journal}{J. Phys.\ B} \textbf{\bibinfo{volume}{37}},
  \bibinfo{pages}{4193} (\bibinfo{year}{2004}).

\bibitem[{\citenamefont{Safronova and Johnson}(2007)}]{review07}
\bibinfo{author}{\bibfnamefont{M.~S.} \bibnamefont{Safronova}}
  \bibnamefont{and} \bibinfo{author}{\bibfnamefont{W.~R.}
  \bibnamefont{Johnson}}, \bibinfo{journal}{Adv. At. Mol., Opt. Phys.}
  \textbf{\bibinfo{volume}{55}}, \bibinfo{pages}{191} (\bibinfo{year}{2007}).

\bibitem[{\citenamefont{Johnson et~al.}(1988)\citenamefont{Johnson, Blundell,
  and Sapirstein}}]{Bspline}
\bibinfo{author}{\bibfnamefont{W.~R.} \bibnamefont{Johnson}},
  \bibinfo{author}{\bibfnamefont{S.~A.} \bibnamefont{Blundell}},
  \bibnamefont{and}
  \bibinfo{author}{\bibfnamefont{J.}~\bibnamefont{Sapirstein}},
  \bibinfo{journal}{Phys.\ Rev.\ A} \textbf{\bibinfo{volume}{37}},
  \bibinfo{pages}{307} (\bibinfo{year}{1988}).

\bibitem[{\citenamefont{Johnon et~al.}(1996)\citenamefont{Johnon, Liu, and
  Sapirstein}}]{adndt-96}
\bibinfo{author}{\bibfnamefont{W.~R.} \bibnamefont{Johnon}},
  \bibinfo{author}{\bibfnamefont{Z.~W.} \bibnamefont{Liu}}, \bibnamefont{and}
  \bibinfo{author}{\bibfnamefont{J.}~\bibnamefont{Sapirstein}},
  \bibinfo{journal}{At.\ Data and Nucl.\ Data Tables}
  \textbf{\bibinfo{volume}{64}}, \bibinfo{pages}{279} (\bibinfo{year}{1996}).

\bibitem[{\citenamefont{Safronova et~al.}(1999)\citenamefont{Safronova,
  Johnson, and Derevianko}}]{mar-pol-99}
\bibinfo{author}{\bibfnamefont{M.~S.} \bibnamefont{Safronova}},
  \bibinfo{author}{\bibfnamefont{W.~R.} \bibnamefont{Johnson}},
  \bibnamefont{and}
  \bibinfo{author}{\bibfnamefont{A.}~\bibnamefont{Derevianko}},
  \bibinfo{journal}{Phys.\ Rev.\ A} \textbf{\bibinfo{volume}{60}},
  \bibinfo{pages}{4476} (\bibinfo{year}{1999}).

\bibitem[{\citenamefont{Safronova and Safronova}(2011)}]{SafSaf11rb}
\bibinfo{author}{\bibfnamefont{M.~S.} \bibnamefont{Safronova}}
  \bibnamefont{and} \bibinfo{author}{\bibfnamefont{U.~I.}
  \bibnamefont{Safronova}}, \bibinfo{journal}{Phys. Rev. A}
  \textbf{\bibinfo{volume}{83}}, \bibinfo{pages}{052508}
  (\bibinfo{year}{2011}).

\bibitem[{\citenamefont{Safronova and Safronova}(2012)}]{SafSaf12}
\bibinfo{author}{\bibfnamefont{M.~S.} \bibnamefont{Safronova}}
  \bibnamefont{and} \bibinfo{author}{\bibfnamefont{U.~I.}
  \bibnamefont{Safronova}}, \bibinfo{journal}{Phys.\ Rev.\ A}
  \textbf{\bibinfo{volume}{85}}, \bibinfo{pages}{022504}
  (\bibinfo{year}{2012}).

\bibitem[{\citenamefont{Safronova and Safronova}(2013)}]{SafSaf13}
\bibinfo{author}{\bibfnamefont{M.~S.} \bibnamefont{Safronova}}
  \bibnamefont{and} \bibinfo{author}{\bibfnamefont{U.~I.}
  \bibnamefont{Safronova}}, \bibinfo{journal}{Phys.\ Rev.\ A}
  \textbf{\bibinfo{volume}{87}}, \bibinfo{pages}{032501}
  (\bibinfo{year}{2013}).

\bibitem[{Ral()}]{RalKraRea11}
\bibinfo{note}{Yu.~Ralchenko, A.~Kramida, J.~Reader, and NIST ASD Team (2011).
  NIST Atomic Spectra Database (version 4.1), http://physics.nist.gov/asd.
  National Institute of Standards and Technology, Gaithersburg, MD.}

\bibitem[{\citenamefont{Johnson et~al.}(1995)\citenamefont{Johnson, Plante, and
  Sapirstein}}]{multipol}
\bibinfo{author}{\bibfnamefont{W.~R.} \bibnamefont{Johnson}},
  \bibinfo{author}{\bibfnamefont{D.~R.} \bibnamefont{Plante}},
  \bibnamefont{and}
  \bibinfo{author}{\bibfnamefont{J.}~\bibnamefont{Sapirstein}},
  \bibinfo{journal}{Adv.\ Atom.\ Mol.\ Opt.\ Phys.}
  \textbf{\bibinfo{volume}{35}}, \bibinfo{pages}{255} (\bibinfo{year}{1995}).

\bibitem[{\citenamefont{Mitroy et~al.}(2010)\citenamefont{Mitroy, Safronova,
  and Clark}}]{MitSafCla10}
\bibinfo{author}{\bibfnamefont{J.}~\bibnamefont{Mitroy}},
  \bibinfo{author}{\bibfnamefont{M.~S.} \bibnamefont{Safronova}},
  \bibnamefont{and} \bibinfo{author}{\bibfnamefont{C.~W.} \bibnamefont{Clark}},
  \bibinfo{journal}{J. Phys. B} \textbf{\bibinfo{volume}{43}},
  \bibinfo{pages}{202001} (\bibinfo{year}{2010}).

\bibitem[{\citenamefont{Keele et~al.}(2012)\citenamefont{Keele, Smith, Lundeen,
  and Fehrenbach}}]{18192EL}
\bibinfo{author}{\bibfnamefont{J.~A.} \bibnamefont{Keele}},
  \bibinfo{author}{\bibfnamefont{C.~S.} \bibnamefont{Smith}},
  \bibinfo{author}{\bibfnamefont{S.~R.} \bibnamefont{Lundeen}},
  \bibnamefont{and} \bibinfo{author}{\bibfnamefont{C.~W.}
  \bibnamefont{Fehrenbach}}, \bibinfo{journal}{Phys. Rev. A}
  \textbf{\bibinfo{volume}{85}}, \bibinfo{pages}{064502}
  (\bibinfo{year}{2012}).

\bibitem[{\citenamefont{Itano}(2006)}]{I06}
\bibinfo{author}{\bibfnamefont{W.~M.} \bibnamefont{Itano}},
  \bibinfo{journal}{Phys. Rev. A} \textbf{\bibinfo{volume}{73}},
  \bibinfo{pages}{022510} (\bibinfo{year}{2006}).

\bibitem[{\citenamefont{{Jiang} et~al.}(2008)\citenamefont{{Jiang}, {Arora},
  and {Safronova}}}]{JiaAroSaf08}
\bibinfo{author}{\bibfnamefont{D.}~\bibnamefont{{Jiang}}},
  \bibinfo{author}{\bibfnamefont{B.}~\bibnamefont{{Arora}}}, \bibnamefont{and}
  \bibinfo{author}{\bibfnamefont{M.~S.} \bibnamefont{{Safronova}}},
  \bibinfo{journal}{\pra} \textbf{\bibinfo{volume}{78}}, \bibinfo{eid}{022514}
  (\bibinfo{year}{2008}).

\bibitem[{\citenamefont{{Roos} et~al.}(2006)\citenamefont{{Roos}, {Chwalla},
  {Kim}, {Riebe}, and {Blatt}}}]{quad2}
\bibinfo{author}{\bibfnamefont{C.~F.} \bibnamefont{{Roos}}},
  \bibinfo{author}{\bibfnamefont{M.}~\bibnamefont{{Chwalla}}},
  \bibinfo{author}{\bibfnamefont{K.}~\bibnamefont{{Kim}}},
  \bibinfo{author}{\bibfnamefont{M.}~\bibnamefont{{Riebe}}}, \bibnamefont{and}
  \bibinfo{author}{\bibfnamefont{R.}~\bibnamefont{{Blatt}}},
  \bibinfo{journal}{Nature} \textbf{\bibinfo{volume}{443}},
  \bibinfo{pages}{316} (\bibinfo{year}{2006}).

\end{thebibliography}
\end{document}